\documentclass[sigconf]{acmart}
\AtBeginDocument{%
  }

\copyrightyear{2026}
\acmYear{2026}
\setcopyright{cc}
\setcctype{by}
\acmConference[DIS '26]{Designing Interactive Systems Conference}{June 13--17, 2026}{Singapore, Singapore}
\acmBooktitle{Designing Interactive Systems Conference (DIS '26), June 13--17, 2026, Singapore, Singapore}
\acmDOI{10.1145/3800645.3812899}
\acmISBN{979-8-4007-2563-0/2026/06}

\usepackage{makecell}
\usepackage{subfig}
\usepackage{tikz}
\usepackage{enumitem}
\usepackage{cancel}
\usetikzlibrary{arrows.meta, positioning, decorations.pathreplacing, shapes.geometric, calc, fit, backgrounds}
\usepackage{xcolor}
\usepackage{fontawesome5}

\definecolor{workflowblue}{RGB}{59, 130, 246}
\definecolor{normangreen}{RGB}{34, 197, 94}
\definecolor{executiongulf}{RGB}{239, 68, 68}
\definecolor{evaluationgulf}{RGB}{249, 115, 22}
\definecolor{temporalbarrier}{RGB}{139, 92, 246}
\definecolor{lightexecution}{RGB}{254, 226, 226}
\definecolor{lightevaluation}{RGB}{255, 237, 213}

\begin{document}

\title{MyoInteract: A Framework for Fast Prototyping of Biomechanical HCI Tasks using Reinforcement Learning}

\author{Ankit Bhattarai}
\authornote{Equal first author contribution.}
\orcid{0009-0007-8233-4344} 
\affiliation{%
  \institution{University of Cambridge}
  \city{Cambridge}
  \country{United Kingdom}}
\email{ab2731@cam.ac.uk }

\author{Hannah Selder}
\authornotemark[1]
\orcid{0009-0008-7049-1630} 
\affiliation{%
  \institution{Center for Scalable Data Analytics and Artificial Intelligence (ScaDS.AI) Dresden/Leipzig, Leipzig University}
  \city{Leipzig}
  \country{Germany}
}
\email{hannah.selder@uni-leipzig.de}

\author{Florian Fischer}
\authornotemark[1]
\orcid{0000-0001-7530-6838} 
\affiliation{%
  \institution{University of Cambridge}
  \city{Cambridge}
  \country{United Kingdom}}
\email{fjf33@cam.ac.uk}

\author{Arthur Fleig}
\authornote{Equal last author contribution.}
\orcid{0000-0003-4987-7308} 
\affiliation{%
  \institution{Center for Scalable Data Analytics and Artificial Intelligence (ScaDS.AI) Dresden/Leipzig, Leipzig University}
  \city{Leipzig}
  \country{Germany}
}
\email{arthur.fleig@uni-leipzig.de}

\author{Per Ola Kristensson}
\authornotemark[2]
\orcid{0000-0002-7139-871X} 
\affiliation{%
  \institution{University of Cambridge}
  \city{Cambridge}
  \country{United Kingdom}}
\email{pok21@cam.ac.uk}

\begin{abstract}
Reinforcement learning (RL)-based biomechanical simulations have the potential to revolutionize HCI research and interaction design, but currently lack usability and interpretability. Using the Human Action Cycle as a design lens, we identify key limitations of biomechanical RL frameworks and 
develop MyoInteract, a novel framework for fast prototyping of biomechanical HCI tasks. MyoInteract allows designers to setup tasks, user models, and training parameters from an easy-to-use GUI within minutes. It trains and evaluates muscle-actuated simulated users within minutes, reducing training times by up to 98\%. A workshop study with 12 interaction designers revealed that MyoInteract allowed novices in biomechanical RL to successfully setup, train, and assess goal-directed user movements within a single session. By transforming biomechanical RL from a days-long expert task into an accessible hour-long workflow, this work significantly lowers barriers to entry and accelerates iteration cycles in HCI biomechanics research.

\end{abstract}



\maketitle

\section{Introduction}

\begin{figure*}[t]
    \centering
    \includegraphics[width=\textwidth]{teaser_2.png}
    \caption{The GUI of \textit{MyoInteract}, a framework that allows
    HCI researchers to setup, train, monitor, and evaluate biomechanical simulations of interactive user behavior. Users can (1) choose from a list of pre-defined task configurations for different HCI contexts (Mobile Touch, Public Display, Augmented Reality) and (2) adjust them, or define entirely new interaction tasks that involve sequences of target acquisition and selection. They can double-check the rendered interaction environment and (3) adjust the proposed reward function. 
    Enabled by a new GPU-accelerated backend, training with \textit{MyoInteract} (4) takes less than one hour (previously: 12--48 hours) and (5) can be continuously monitored via success metrics.
    (6) Videos enable qualitative inspection of learned behavior. Advanced mode (not shown) offers the option of setting additional parameters relating to task setup, the biomechanical model, and the RL training procedure. %
    }
    \Description{A screenshot of the MyoInteract GUI. The GUI is divided into two sections, with numbered blue circles highlighting key features.
    The left section has a drop-down menu labeled "Pre-saved configurations" with step (1) "Load configuration for the desired interaction task". Below this, there is step (2) "Setup your own task based on the loaded configuration or from scratch by defining a schedule of target reaching and selection tasks". This section also includes task parameters and target setup options, such as a field for setting the number of targets and options for selecting the target type (box or sphere). Additionally, sliders and input fields are present for adjusting box settings, including depth, position, and vertical position.
    The right section features a rendered environment view and step (3) "Adjust predefined reward function" with a table for setting reward weights. A clock icon indicates that training takes 10-60 minutes for step (4) "Run training". Below this, there is step (5) "Monitor success metrics" with two graphs displaying success metrics, including a 3D scatter plot and a line chart. Finally, step (6) "Observe videos of learned behavior" is represented by two video windows showing final policy views from different angles.}
    \label{fig:teaser}
\end{figure*}

Simulations are becoming central to HCI, allowing researchers to ask controlled ``what-if'' questions, test designs at scale for diverse user groups, and probe our understanding of interactive systems. As \citeauthor{murray-smith_what_2022}~\cite{murray-smith_what_2022} argue, the ability to match user behavior with a generative model is a strong test of understanding, making design and engineering more predictable. \textit{Biomechanical Reinforcement Learning} (RL), which trains simulated users to control realistic musculoskeletal models, is a particularly promising direction for producing such generative models.

By simulating interaction at the level of muscles and body dynamics, biomechanical RL can predict motion trajectories and physical effort for complex tasks~\cite{bachynskyi2015performance, caggiano2023myodex, nakada_deep_2018, ikkala_breathing_2022}. Previous work has validated that these simulated users reproduce established motor patterns, such as the bell-shaped velocity profiles of aimed reaching~\cite{fischer2021reinforcement, moon_real-time_2024}. Crucially for HCI, synthesized trajectories align with Fitts' Law~\cite{ikkala_breathing_2022, fischer2021reinforcement, moon_real-time_2024} and the Two-Thirds Power Law~\cite{fischer2021reinforcement, LACQUANITI1983115}, suggesting that simulated users can reliably predict plausible aimed body movements. Consequently, researchers have begun applying these models to button pressing~\cite{ikkala_breathing_2022}, keyboard typing~\cite{Hetzel2021}, joysticks~\cite{ikkala_breathing_2022}, smartphone use~\cite{miazga2025increasing}, and virtual pointing techniques~\cite{klar_simulating_2023, moon_amortized_2023}, as well as using them to guide the ergonomic design of VR/XR interfaces~\cite{fischer_sim2vr_2024, evangelista2021xrgonomics, Miki2025Enhancing} and adaptive systems~\cite{li2024nicer, li2025alphapig, peternel2019selective}.
In the future, such simulations could enable early-stage prototyping---testing layouts, comparing movement strategies, or exploring accessibility for users with diverse motor abilities---without the cost of human studies.
Some visionary examples include a replicated prototype study for a VR game using simulated users~\cite{fischer_sim2vr_2024}, and~\citet{moon_real-time_2024}, who have been able to train a state-of-the-art target inference classifier solely from simulation data.

Despite these promising advances, biomechanical RL currently remains a specialized technique restricted to experts, rather than a practical prototyping tool for the wider HCI community. The expertise required to configure a simulation is prohibitive as setting up a single experiment involves writing model and task specifications, scripting in Python, and designing custom reward functions. Furthermore, the feedback loop is excruciatingly slow: training a policy typically takes 12 hours to several days~\cite{ikkala_breathing_2022, selder2025demystifying, fischer_sim2vr_2024, moon_real-time_2024}, with little visibility during the training process into whether the configuration is succeeding. If the training produces implausible behavior, diagnosis means restarting the training.

\begin{figure*}
    \centering
    \includegraphics[width=\linewidth]{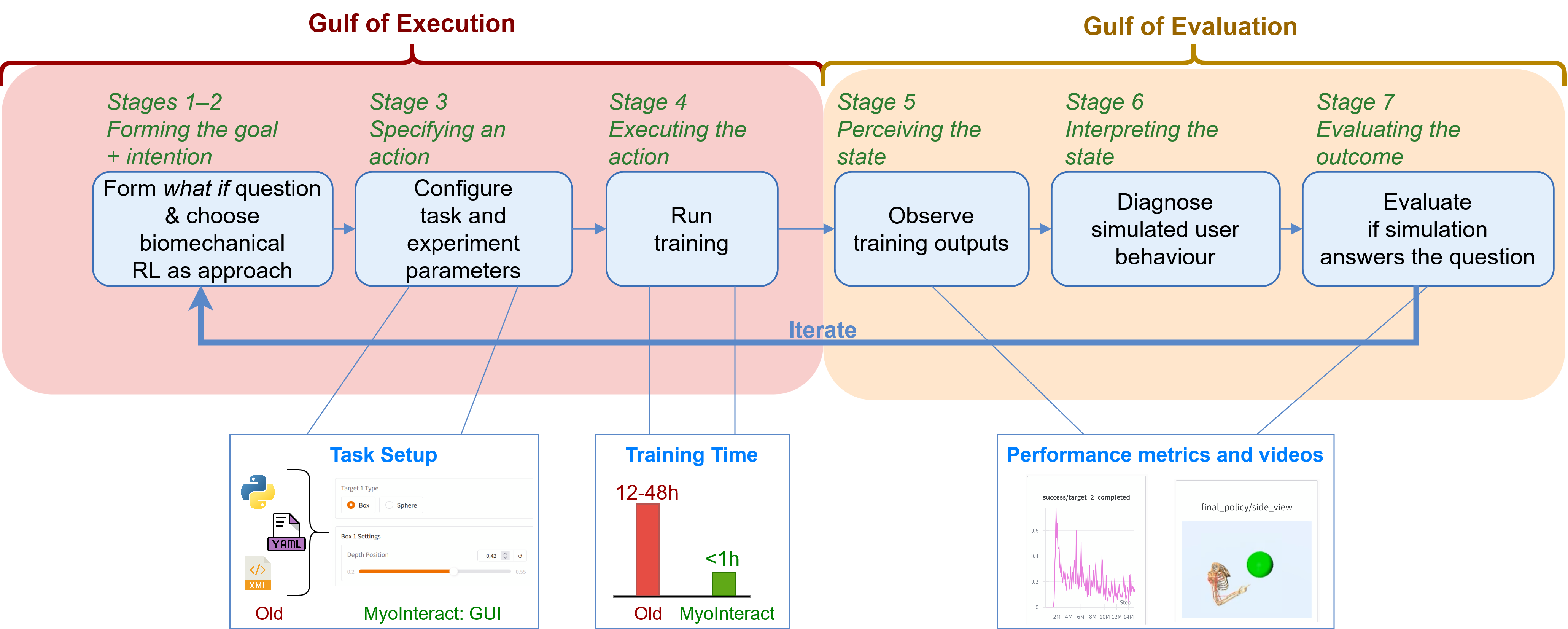}
    \caption{Workflow of the biomechanical RL simulations approach analyzed using Norman's seven stages of action~\cite{Norman2002-da}, 
    illustrating barriers in both the Gulf of Execution (task specification) and the Gulf of Evaluation (training feedback). The bottom panels show how MyoInteract addresses each barrier: a GUI replaces manual XML/Python/YAML editing, GPU acceleration compresses training from 12–48 hours to under one hour, and a 
    suite of performance metrics and visualizations 
    replaces the minimal outputs of prior systems (which provided only aggregate metrics such as mean reward and episode length). These features significantly shorten the action cycle, transforming biomechanical RL simulation from a days-long expert endeavor into an accessible, hour-scale workflow.}
    \label{fig:norman_workflow}
    \Description{A flowchart illustrating the biomechanical reinforcement learning (RL) simulations approach, analyzed using Norman's seven stages of action.
    The flowchart is divided into two main sections: the "Gulf of Execution" and the "Gulf of Evaluation". The Gulf of Execution outlines the initial stages of the workflow, starting with stages 1-2, which focus on forming the goal and intention by formulating what-if questions and selecting biomechanical RL as the approach. The next stage, stage 3, involves specifying an action by configuring task and experiment parameters. This is followed by stage 4, executing the action by running training. The Gulf of Evaluation outlines the subsequent stages, starting with stage 5, perceiving the state by observing training outputs. The next stage, stage 6, involves interpreting the state by diagnosing simulated user behavior. Finally, stage 7 involves evaluating the outcome by assessing whether the simulation answers the question. The flowchart also highlights several key components, including a task setup interface with options for target type, such as box or sphere, and adjustable box settings. A bar graph compares training times, showing that MyoInteract significantly reduces training time from 12-48 hours to less than 1 hour. Additionally, a section displays new performance metrics and videos, featuring a line graph and a final policy view. The workflow involves iterating between the Gulf of Execution and the Gulf of Evaluation to refine and improve the simulation.}
\end{figure*}

Viewed through Norman's action cycle~\cite{Norman2002-da}, prototyping biomechanical HCI tasks %
suffers from severe \textit{Gulfs of Execution and Evaluation} (see Figure~\ref{fig:norman_workflow}).
Consider a designer exploring widget placement in Mixed Reality to minimize physical effort. %
Testing different configurations \textit{in silico} faces a \textit{Gulf of Execution}: the designer cannot simply place a target; they need to modify XML files and Python scripts to define the task, adjust RL parameters to ensure training succeeds, and understand low-level details of the simulation pipeline~\cite{ikkala_breathing_2022}. Once training begins, the process becomes a black box, creating a \textit{Gulf of Evaluation}. The system offers no diagnostic insight into why a simulated user behaves in a certain way, or how this behavior emerged. 
Currently, no systematic diagnosis and analysis tools exist to assess and compare the infinite number of possible design choices. %
Combined with day-long training times, these gulfs render iterative prototyping impossible. As long as setting up a simulation takes longer than running a human study, biomechanical RL cannot serve its purpose as a design tool. %
This paper asks: \textit{What if these barriers could be collapsed, transforming biomechanical RL from a specialized research capability into an accessible, practical instrument for HCI prototyping?}

We present \textit{MyoInteract}, a framework designed to address three design goals: reducing the gulf of execution, reducing the gulf of evaluation, and collapsing the temporal barrier of biomechanical user simulations. %
\textit{MyoInteract} leverages MuJoCo-MJX~\cite{ mujoco_mjx},
 a GPU-accelerated implementation of the MuJoCo physics engine~\cite{todorov_mujoco_12} based on the JAX framework~\cite{jax2018github} %
that for the first time enables training and running  simulations entirely on the GPU. 
Utilizing this JAX-based parallelization and acceleration, we re-implement movement-based interaction tasks, focusing on sequential target acquisition and selection tasks, for which biomechanical RL simulations have been extensively validated~\cite{fischer2021reinforcement, ikkala_breathing_2022, moon_real-time_2024, fischer_sim2vr_2024, miazga2025increasing}. 
This 
reduces the training time by up to 98\% from days to minutes while preserving biomechanically plausible movement patterns. In addition, \textit{MyoInteract} offers a graphical interface (Figure~\ref{fig:teaser}) that allows users to configure tasks and parameters without extensive coding or RL expertise. Finally, it provides several HCI-relevant metrics %
that surface immediate and comprehensive insights into learned behavior during training. 
We report on a hands-on workshop with 12 HCI researchers and practitioners. 
We find that by collapsing the time and expertise barriers, \textit{MyoInteract} transforms a days-long expert endeavor into an accessible, hour-scale workflow,  shifting biomechanical simulation into an interactive, iterative activity for the first time.

This work \textbf{contributes}:
\begin{enumerate}
    \item \textbf{\textit{MyoInteract}}, %
    a framework for fast prototyping of biomechanical target selection HCI tasks that enables the configuration, simulation, and evaluation of interaction movements in less than an hour. The open-source code is available at \textcolor{blue}{\url{https://github.com/MyoInteract/MyoInteract}}.%
    \item \textbf{Findings} from a workshop with 12 HCI researchers, %
    yielding empirical insights that the framework allows biomechanical RL novices to successfully deploy biomechanical user simulations, and providing suggestions for designing biomechanical user simulation tools.
\end{enumerate}
The remainder of the paper situates this work in related literature (Section~\ref{sec:related-work}), illustrates our design rationale (Section \ref{sec:design_rationale}), 
describes the framework and interface (Section~\ref{sec:our-methods}), shows demonstrations and performance analyses (Section~\ref{sec:demonstrations}), presents findings from the workshop evaluation (Section~\ref{sec:workshop-evaluation}), and concludes with a reflection on the opportunities and limitations of biomechanical simulations for HCI and tangible next steps to increase the validity and scope of computational user simulations (Section~\ref{sec:discussion}).

\section{Related Work}\label{sec:related-work}

The idea of simulating users as embodied agents has gained traction in and beyond HCI. Our work builds on the progression from inverse to forward biomechanical models in HCI, the technical foundations that enabled this shift, and the emerging need for accessible computational design tools.

\subsection{From Inverse to Forward Biomechanical Models in HCI}
Within HCI, research on biomechanical simulation has historically relied on \textit{inverse models}, where muscle and joint forces are inferred from recorded motion to estimate fatigue and muscle utilization~\cite{Bachynskyi14, bachynskyi2015performance}. While useful for analyzing existing movements, inverse models cannot predict how users would interact with novel interface designs.

Recently, advances in physics engines like MuJoCo~\cite{todorov_mujoco_12} and specialized toolkits such as MyoSuite~\cite{caggiano2022myosuite} and Hyfydy~\cite{schumacher2023natural} have enabled a shift toward \textit{forward models} that generate behavior directly from muscle activation signals~\cite{ikkala_breathing_2022, klar_simulating_2023, Hetzel2021, moon_real-time_2024}. This shift is critical: forward approaches allow researchers to evaluate interaction techniques and estimate physical effort \emph{before} running user studies. However, forward models require a \emph{controller} to determine muscle activations of the musculoskeletal system on a moment-to-moment basis, a complex high-dimensional control problem.

Reinforcement Learning has emerged as the dominant solution for this control problem. \citeauthor{fischer2021reinforcement}~\cite{fischer2021reinforcement} demonstrated how RL could drive a torque-actuated human arm model to accomplish goal-directed movements, with trajectories that matched established motor control principles. \citeauthor{ikkala_breathing_2022}~\cite{ikkala_breathing_2022} introduced \textit{User-in-the-Box} (UitB), which combined muscle-level dynamics with perceptual feedback to simulate reaching and pointing behavior. Alternative control approaches, such as Model Predictive Control~\cite{klar_simulating_2023}, have also been explored, though RL has proven more scalable for complex, multi-goal tasks. These foundational works have since been extended to VR ergonomics~\cite{fischer_sim2vr_2024} and goal inference~\cite{moon_real-time_2024}, demonstrating the potential of biomechanical RL for analyzing and improving interactive systems.

Beyond HCI, biomechanical models have been used to predict human movements and ergonomic states from muscle control signals~\cite{ackermann2010optimality, maas2013biomechanical}. Biomechanical RL specifically has been adopted by the rehabilitation~\cite{nowakowski2022deep, hong2024biomechanical, coser2024ai}, neuroscience~\cite{simos2025reinforcement, caggiano2023myodex, chiappa2024acquiring, schumacher2023natural}, and robotics communities~\cite{peng2025gait, radosavovic2024real} for movement prediction, prosthesis control, and motor learning research. This widespread adoption across applications confirms RL as a state-of-the-art method for generating plausible muscle-driven behavior.

\subsection{Complexity as a Barrier to Adoption}
While the potential of biomechanical RL is clear, its adoption in HCI is hindered by its complexity. A critical review of the literature reveals that virtually no HCI studies have successfully employed frameworks like \textit{UitB} or \textit{MyoSuite} ``out of the box''. Instead, successful implementation typically requires (extensive) modification of reward functions, biomechanical models, and training procedures. \citeauthor{Miki2025Enhancing}~\cite{Miki2025Enhancing} notably reported ``poor generalizability of the [UitB] simulator'', underscoring that even minor changes, such as altering arm length, require tedious manual parameter tuning and retraining.

Consequently, a significant portion of recent research has been dedicated solely to making these simulations work reliably. \citeauthor{selder2025demystifying}~\cite{selder2025demystifying,selder2025whatmakes} investigated the fragility of reward structures, deriving design guidelines to prevent simulation failures. Similarly, specific learning curricula are required to successfully train simulated users for dexterous tasks~\cite{miazga2025increasing}. Further work has explored how motor constraints~\cite{charaja_generating_2024} and noise~\cite{chiappa2023latent} influence learning efficiency. These works illustrate that biomechanical RL is currently less of a design tool and more of a technical research challenge.

\subsection{The Wider Landscape of User Simulation}
Parallel to biomechanics, a wider movement in HCI explores how simulation can model user behavior more generally. Frameworks based on Computational Rationality~\cite{oulasvirta_computational_2022,liao25affordance} and Active Inference~\cite{murraysmith25activeinference} use optimization to explain how users balance cognitive and motor costs. RL has also been employed to model adaptive decision-making in non-biomechanical contexts, such as touchscreen typing~\cite{shi_crtypist_2024} and adaptive interface control~\cite{langerak_marlui_2024}. More recently, Large Multimodal Models have been used as agentic user proxies~\cite{zhang2025socioverse} or to infer preferences from unstructured data~\cite{Shaikh2025}.

These developments have broadened the scope of user simulation from the physical to the cognitive and social levels. 
However, these cognitive and social intents must ultimately be mediated by the physical body.
Currently, we lack a framework for the \textit{physical body} that supports \textit{iterative} design, i.e., rapidly exploring how constraints like fatigue and effort shape interaction, rather than waiting days for a single analysis. What is missing is the speed and usability required to transform biomechanical simulation from a technical research method into a practical prototyping instrument. \textit{MyoInteract} addresses this gap by providing GPU-accelerated training, real-time visual diagnostics, and a graphical interface that lets users configure and run biomechanical RL simulations without deep technical expertise.

\section{Design Goals}\label{sec:design_rationale}

Our design process was guided by the goal of making biomechanical simulation accessible to HCI researchers without sacrificing the fidelity required for valid user modeling. Using Norman's Action Cycle~\cite{Norman2002-da} as a diagnostic lens, we derived three aspirational \textbf{design goals} that shaped our development of \textit{MyoInteract}:
\begin{enumerate}[label=DG\arabic*]
\item\label{dg-execution} \textbf{Reduce the Gulf of Execution:} Eliminate the need for XML/YAML/Python coding by raising the level of abstraction, allowing researchers to specify \textit{what} they want to test rather than \textit{how} to implement it.
\item\label{dg-evaluation} \textbf{Reduce the Gulf of Evaluation:} Make the opaque process of RL training more observable by surfacing diagnostic information in real-time, helping users distinguish task design issues from configuration errors and training failures. 
\item\label{dg-speed} \textbf{Collapse the Temporal Barrier:} Compress the long training times that compound both gulfs to enable iterative exploration within a single working session.
\end{enumerate}
While our current implementation focuses on sequential pointing and selection tasks, the principles we derive are intended to generalize to broader biomechanical simulation workflows.

\subsection{Design Principles and System Evolution}\label{sec:design_principles}

We developed \textit{MyoInteract} iteratively, guided by six principles that emerged as we moved from internal prototypes to the final system. These principles address the barriers identified above, scoped to the domain of upper-body pointing and selection tasks.

\paragraph{DP1: Enable Rapid Iteration through Speed ($\rightarrow$ \ref{dg-speed})}\label{dp1}
For simulation to support design exploration, feedback loops must be short enough to permit multiple iterations within a session. In prior work, researchers had to commit to a single configuration and wait hours or days to assess viability. Drawing on recent advances in GPU-accelerated RL environments~\cite{thibault2024learning, zakka2025mujoco}, we prioritized reducing training time not merely for efficiency, but to fundamentally change the character of the workflow from batch analysis to interactive exploration. By compressing training to minutes (Section~\ref{sec:mjx_dp1}), users can test alternative designs, reward functions, and task variants iteratively.

\paragraph{DP2: Enable Composability through Task Decomposition ($\rightarrow$ \ref{dg-execution})}\label{dp2}
In prior frameworks like~\cite{ikkala_breathing_2022} and our early prototypes, each interaction scenario was implemented as a monolithic unit, requiring new XML files and custom Python logic. This high ``viscosity''~\cite{GREEN1996131} discouraged experimentation with complex sequences. We addressed this by decomposing tasks into sequential combinations of pointing and button-pressing primitives. Users now specify tasks through high-level configuration instead of code, and the system automatically generates the corresponding simulation logic (Section~\ref{sec:combinatorial_task_decomposition}). This abstraction reduces setup time and lowers the expertise barrier.

\paragraph{DP3: Make Configuration Visible and Manipulable ($\rightarrow$ \ref{dg-execution})}\label{dp3}
While configuration files reduce code changes, they are prone to silent errors: users may be unsure which parameters are valid, how they affect behavior, or which defaults apply without consulting documentation. Guided by Norman's principle of Visibility~\cite{Norman2002-da}, %
we replaced text-based configuration with a graphical interface that exposes available parameters, displays current and default values, and enforces valid ranges through interactive controls (Section~\ref{sec:GUI}). Users can also preview target layouts spatially before training, reducing the likelihood of misconfigured tasks.

\paragraph{DP4: Provide Multi-Level Feedback for Diagnosis ($\rightarrow$ \ref{dg-evaluation})}\label{dp4}
RL failures are notoriously opaque: a success rate that plateaus at 70\% may reflect unreachable targets, misspecified rewards, or insufficient training. To support diagnosis, we provide feedback at multiple stages. 
Pre-training, targets are specified using numerical parameters that define their 3D position. To visualize how these parameters affect the spatial layout, we augment the GUI with a rendering feature that allows users to preview the task environment. %
During training, real-time dashboards (rather than upon completion) decompose aggregate metrics (e.g., per-target success rates, individual reward components). Originally developed as internal debugging tools, these proved essential for identifying bottlenecks (e.g., ``the fourth target in a sequence is unreachable'') that aggregate scores obscure (Section~\ref{sec:metrics}).
Post-training, automatically generated videos reveal qualitative issues such as jittery motion or unnatural postures, which numerical metrics alone cannot easily surface.

\paragraph{DP5: Encode Domain Constraints to Prevent Errors ($\rightarrow$ \ref{dg-execution})}\label{dp5}
Biomechanical RL configurations are fragile; small parameter errors can invalidate results. Through iterative development enabled by faster training cycles, we identified common failure modes and encoded safeguards into the interface. For example, sliders enforce valid parameter ranges, target placements are constrained to plausible workspace regions, and the system recommends training durations based on task complexity. These constraints reduce the likelihood of wasted runs, particularly for novice users.

\paragraph{DP6: Support Progressive Disclosure ($\rightarrow$ \ref{dg-execution})}\label{dp6}
Early GUI iterations exposed all parameters simultaneously, overwhelming users, especially novices, who struggled to distinguish essential settings from advanced options. Inspired by progressive disclosure principles~\cite{lidwell2010universal}, we split the interface into a ``Simple Mode'' for task setup and an ``Advanced Mode'' for precise control. This structure supports a natural workflow: users begin with coarse task design and reveal additional control only when needed, balancing accessibility with expert capability.

\section{MyoInteract: A GPU-accelerated Interaction Simulation Framework}\label{sec:our-methods}
In this section, we describe how the design principles outlined in Section~\ref{sec:design_rationale} are realized in \textit{MyoInteract}, a GPU-accelerated framework for biomechanical user simulation. Our design choices prioritize speed, composability, and observability. 

\subsection{System Architecture: Speed through Parallelization (DP1)}
\label{sec:mjx_dp1}

To enable rapid iteration (DP1), we built \textit{MyoInteract} on MuJoCo-MJX~\cite{mujoco_mjx}%
, a GPU-accelerated re-implementation of the MuJoCo physics engine~\cite{todorov_mujoco_12} in JAX~\cite{jax2018github}. Unlike CPU-based simulators, MJX enables thousands of parallel environments on GPUs/TPUs, dramatically increasing training speed. Policy learning uses PPO~\cite{Schulman_17_PPO} implemented in Brax~\cite{brax2021github} (hyperparameters in Appendix~\ref{sec:hyperparams}), enabling end-to-end GPU execution without CPU-GPU transfer bottlenecks.

While recent work has leveraged MJX for humanoid locomotion~\cite{thibault2024learning, zakka2025mujoco} and pose estimation~\cite{firouzabadi2024biomechanical}, to our knowledge, \textit{MyoInteract} is the first framework to apply GPU-accelerated biomechanical simulation specifically to interactive HCI tasks. This architectural choice is the technical foundation for DP1, providing the computational throughput necessary to shift the workflow toward interactive exploration.

The framework integrates with the open-source MyoSuite library~\cite{caggiano2022myosuite}, ensuring compatibility with a broad range of musculoskeletal models~\cite{walia2025myoback, wang2022myosim} and emerging extensions such as fatigue models, assistive devices~\cite{tan2025myoassist}, and sample-efficient RL methods~\cite{chiappa2023latent, schumacher_dep-rl_2022}.

\subsection{Composable Interaction Modeling (DP2)}
To realize the goal of task composability (DP2), \textit{MyoInteract} provides a unified environment where users can configure the biomechanical model, the task, and the information flow between them (observations) without low-level coding.
In the following, we describe these components conceptually, while implementation details, including parameter values, are provided in the appendix.

\subsubsection{Biomechanical Model}\label{sec:biom-model-details}
\textit{MyoInteract} is designed to support a range of musculoskeletal models. As the default for pointing tasks, we employ an adaptation of the MoBL-ARMS model~\cite{saul2015benchmarking}, the same high-fidelity representation of the upper extremity with 26 muscles and fixed fingers provided in the UitB framework~\cite{ikkala_breathing_2022}. However, users can toggle between alternative models via the GUI, including the MyoArm\footnote{\url{https://github.com/MyoHub/myo_sim}} and a full-hand extension of the MoBL-ARMS model.

To capture the motor variability underlying the speed-accuracy trade-off~\cite{harris1998signal, Van_Beers2004-mp}, we implement both signal-dependent and constant motor noise. Given the ongoing debate regarding the optimal noise structure for simulation realism~\cite{fischer2021reinforcement, charaja_generating_2024}, we expose these parameters as configurable options in the GUI, allowing researchers to empirically test their effect on plausibility. 
Finally, to ensure robust policies that generalize across postures rather than overfitting to a single starting position, the system randomizes initial joint states and muscle activations at the start of each episode. %

\subsubsection{Task Composability}\label{sec:combinatorial_task_decomposition}
Many mid-air HCI scenarios, such as selecting items from AR menus, pressing virtual buttons on public displays, or tapping keys on mid-air keyboards, are fundamentally based on target selection, which is thus, along with manipulation, considered the basis for 3D interaction \cite{alma991006699997403606}. 
We 
decompose interaction into two \textbf{interaction primitives} (DP2) that form the foundational building blocks for the majority of mid-air target selection HCI tasks: 
\begin{itemize}
\item \textbf{Pointing:} The user must move the end-effector (fingertip) into a target sphere and maintain it for a specified dwell time. Targets are parametrized by size, 3D location, and color.
\item \textbf{Pressing:} The user must apply a specific activation force to a surface. Buttons are parametrized by location, orientation, geometry, color, and force threshold.
\end{itemize}

The power of \textit{MyoInteract} lies in \textbf{chaining} these primitives. Instead of implementing monolithic environments for every new task, researchers can define complex workflows as sequences of primitives -- without modifying a single line of XML or Python code. For example, an AR interface can be defined as a sequence alternating between reaching virtual spheres (with 0.5 s dwell) and pressing physical buttons (with 2 N force).
This compositional approach allows for the instantiation of a wide variety of HCI scenarios (Section~\ref{sec:example-demonstrations}). By decoupling task logic from implementation, researchers can focus on the \textit{design} of the interaction sequence rather than the engineering of the simulation.

\subsubsection{Observations}
To control the biomechanical model, the simulated user relies on a state vector combining proprioception (joint angles, velocities, accelerations, muscle activations, fingertip position) and task information (target position/size, sequence progress, dwell timer).
Crucially for design exploration, this observation space is configurable via the GUI (DP3). Users can selectively exclude specific signals to model different user capabilities or constraints, such as masking muscle states to model proprioceptive deficits.

\subsubsection{Reward function}\label{sec:reward-function}
The agent's behavior is shaped by a reward function composed of four weighted terms, where $(\cdot)$ is a placeholder for all relevant function
arguments included in the current system state and muscle control vector:
\begin{itemize}
    \item $f_{\text{distance}}(\cdot)$: a \textbf{distance term} representing the sum of the distances between the fingertip and the current target, as well as between the remaining targets;
    \item $f_{\text{subtask\_bonus}}(\cdot)$: a \textbf{subtask bonus} granted when the current target is successfully reached for the first time;
    \item $f_{\text{completion\_bonus}}(\cdot)$: a \textbf{task completion bonus} awarded upon successful completion of all subtasks;
    \item $f_{\text{effort}}(\cdot)$: an \textbf{effort penalty} that discourages biomechanically implausible or overly strenuous movements. This is achieved by penalizing the squared muscle control values (\textit{neural effort} costs~\cite{berret_evidence_2011, ikkala_breathing_2022, selder2025demystifying}); however, the framework allows replacing this component with alternative effort models.
\end{itemize}
These components are derived from established guidelines~\cite{selder2025demystifying}, and their relative weights determine the emergent movement strategy (e.g., prioritizing speed vs.\ effort).
In \textit{MyoInteract}, these weights are fully exposed in the GUI (DP3), enabling designers to interactively tune this trade-off.

To further simplify training, the system provides suggestions for training duration based on task complexity (e.g., adding 1 million steps per additional target or an additional 0.3 seconds of total dwell time). We iteratively found this heuristic yet sensible baseline to prevent incomplete training runs (supporting DP5).

\subsection{Interface Design: Visibility, Progressive Disclosure, and Guidance (DP3, DP5, DP6)}\label{sec:GUI}
To bridge the gap between high-level interaction design and low-level RL configuration, we developed a graphical interface that replaces manual scripting with direct manipulation. Built on Gradio~\cite{DBLP:journals/corr/abs-1906-02569}, the GUI operationalizes three key design principles: Visibility (DP3), by exposing parameters; Progressive Disclosure (DP6), by layering complexity; and Heuristic Guidance (DP5), by actively assisting with parameter and training decisions.

\begin{figure*}[!ht]
    \centering
    \includegraphics[width=\linewidth]{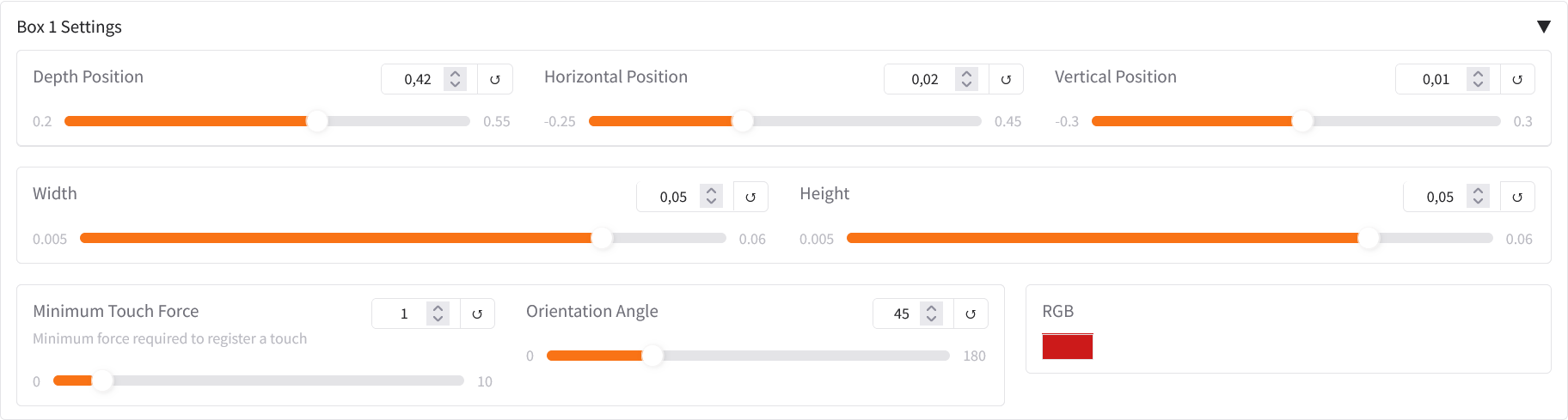}
    \caption{Box target settings in our GUI. Users can adjust position, size, minimum touch force, and orientation angle via sliders, and choose the object color from a palette.}
    \label{fig:gui_box}
    \Description{A screenshot of the box target settings in the GUI, allowing users to customize various parameters. The interface is titled "Box 1 Settings" and features several adjustable settings, including position, size, and additional properties. The position settings include Depth Position, with a slider and input field set to 0.42, Horizontal Position, with a slider and input field set to 0.02, and Vertical Position, with a slider and input field set to 0.01. The size settings include Width, with a slider and input field set to 0.005, and Height, with a slider and input field set to 0.05. Additional settings include Minimum Touch Force, with a slider and input field set to 1, which represents the minimum force required to register a touch, Orientation Angle, with a slider and input field set to 0, and RGB color selection, currently set to red. These settings enable users to fine-tune the box target's properties to suit their needs.}
\end{figure*}

Upon startup, the interface presents users with a direct link to the logging dashboard (Section~\ref{sec:metrics}). Default parameter values and ranges are provided to guide users and reduce the need for parameter tuning (DP5). Configurations can be saved and reloaded to facilitate reuse and reproducibility. To scaffold the learning curve, we structure the interface into two distinct interaction modes:
\paragraph{Simple Mode (Figure~\ref{fig:simple_GUI}).}
Designed for rapid prototyping, this mode exposes the essential parameters for defining interaction logic. Users configure primitives (e.g., spheres, boxes) through direct manipulation controls, adjusting properties such as position, size, and force thresholds via sliders with visual ranges such as those shown in Figure~\ref{fig:gui_box}. A built-in 3D renderer allows users to visually inspect the task environment from lateral and frontal perspectives before training begins, supporting DP4 pre-training. 

Crucially, simple mode also exposes the reward structure. As established in Section~\ref{sec:reward-function}, reward weights dictate the agent's emergent strategy. We do not hide this complexity, but make it transparent: adjusting a weight immediately updates the displayed reward equation, visualizing the mathematical impact of the design choice (DP3). To further aid intuition, tooltips display the theoretical minimum and maximum values for each component based on the current task setup, helping users better calibrate competing objectives, such as speed versus effort.

\paragraph{Advanced Mode (Figure~\ref{fig:advanced_GUI}).} 
Targeting more experienced users, this mode reveals deeper configuration layers (supporting DP6). It provides granular control over three critical dimensions:
\begin{itemize}
\item \textbf{Biomechanical Fidelity:} Users can adjust control timesteps (i.e., the frequency with which the simulated user updates their muscle control), toggle signal-dependent and constant motor noise, and configure whether and how to reset the musculoskeletal model at the end of each episode (e.g., a fixed or random initial pose). 
\item \textbf{Observation Space:} Users can customize the agent's sensory inputs, selectively enabling or disabling specific proprioceptive signals to model user constraints (e.g., specific sensory deficits).
\item \textbf{Training Dynamics:} To prevent incomplete training runs, i.e., too few training steps, the system implements Heuristic Guidance and Domain Constraints (DP5). It automatically calculates a recommended training duration based on task complexity (scaling with target count and dwell time). While novices can rely on this baseline to ensure convergence, experts retain the ability to override it, fine-tune RL hyperparameters (e.g., batch size, number of parallel environments), and manage the frequencies to save the current policy (checkpoint frequency) and to evaluate it (evaluation frequency). Users can also load pre-trained policies to continue training and specify random seeds for reproducibility.
\end{itemize}
This layered approach ensures that while the system remains accessible to HCI researchers without biomechanical RL expertise, it retains the depth required for rigorous computational evaluation.

\subsection{Metrics and Logging (DP4)}\label{sec:metrics}
One of the primary challenges in applying RL to HCI is the ``black box'' nature of training. To address this, \textit{MyoInteract} implements Observability (DP4) by recording a comprehensive suite of metrics designed to facilitate not just monitoring, but active debugging of policy behavior. All metrics are automatically synchronized with Weights \& Biases (WandB)\footnote{WandB (\url{https://wandb.ai/}) is an online machine learning development platform that allows researchers to manage, track, and visualize their training in real time.}, 
providing a centralized dashboard for real-time tracking.

\begin{figure*}[!ht]
    \centering
    \includegraphics[scale=0.45]{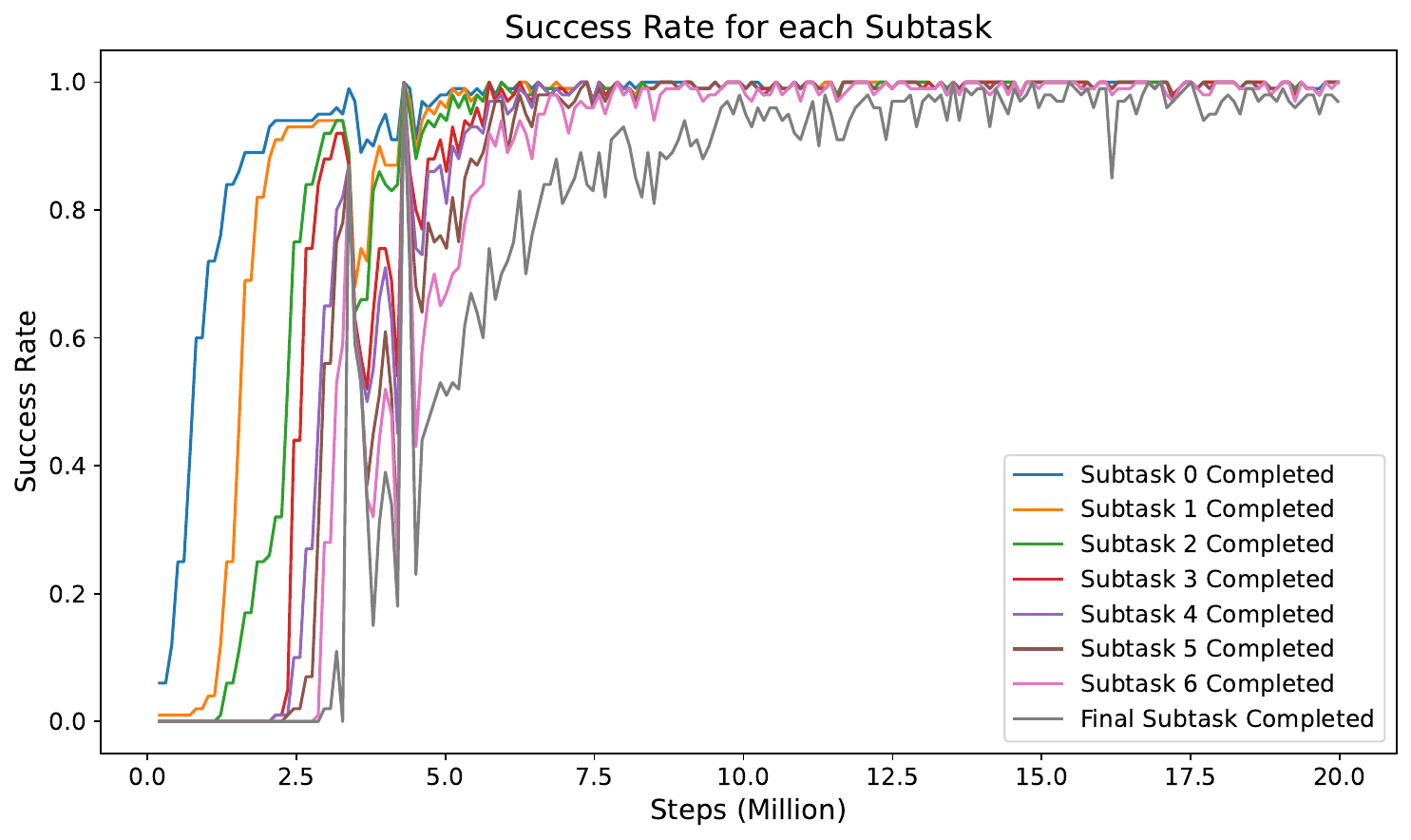}
    \caption{Analysis of subtask completion metrics for each target in the AR task. Note how the success rates of subtasks build upon each other sequentially.}
    \label{fig:task_progression_metrics}
    \Description{A line graph illustrating the success rates of subtask completion metrics for each target in the Augmented Reality (AR) task, displaying the success rates of individual subtasks ranging from 0.0 to 1.0, plotted against the number of steps in millions, with the success rates of subtasks building upon each other sequentially, indicating a cumulative effect where the completion of earlier subtasks contributes to the success of subsequent ones.}
\end{figure*}

\paragraph{Granular Performance Metrics.}
Standard RL metrics (e.g., total reward) are often insufficient for diagnosing failure in complex movement sequences. Therefore, we decompose performance into granular units:
\begin{itemize}
\item \textbf{Sequential Subtask Completion:} Beyond overall success rates, we log completion rates for each individual subtask (Figure~\ref{fig:task_progression_metrics}). This provides direct insights into which parts of the tasks are difficult to achieve.
\item \textbf{Spatial Error Analysis:} To identify biomechanical ``blind spots'', the system generates a 3D visualization showing the success rate for each of the target locations, as well as 2D plots correlating performance with target size. This enables the immediate identification of unreachable workspace areas or precision deficits.
\item \textbf{Terminal Distance Error:} We log the remaining distance to targets at episode termination, distinguishing between ``near misses'' and complete tracking failures.
\end{itemize}

\begin{figure*}[!ht]
    \centering
    \includegraphics[scale=0.5]{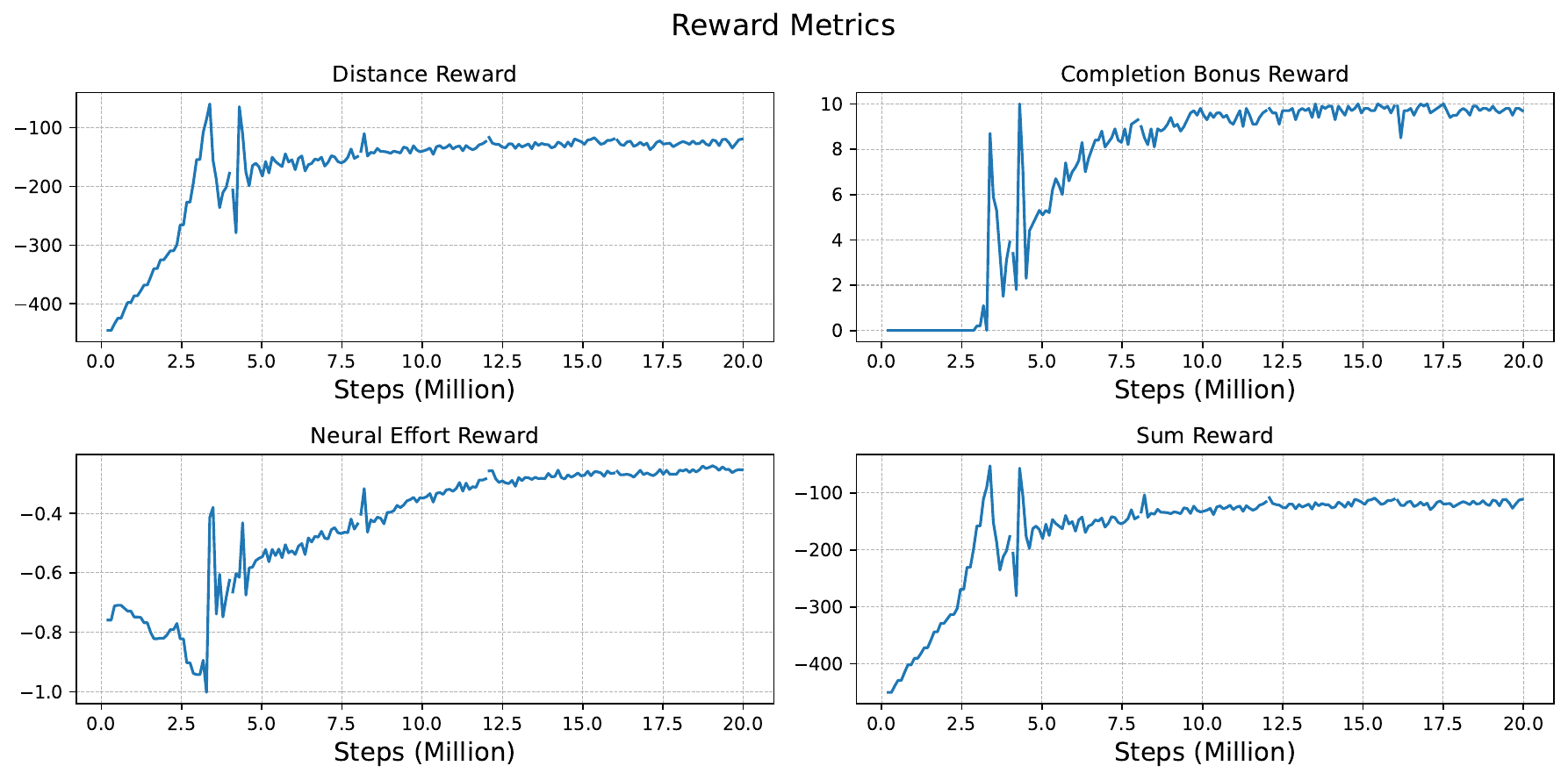}
    \caption{Analysis of various reward components across training for the AR task. In this example, the individual reward components converge after 10-14 million steps, respectively.}
    \label{fig:reward_metrics}
    \Description{A graph showing four line plots for reward metrics, including Distance Reward, Completion Bonus Reward, Neural Effort Reward, and Sum Reward, each displaying a blue line that fluctuates over 20 million steps while increasing. All reward curves are increasing with the distance, completion bonus and sum reward converging around 10 Million steps and neural effort converging at 17.5 Million steps.}
\end{figure*}

\paragraph{Reward Metrics.} %
To make the agent's training incentives more transparent, we log the individual contributions of each reward component (distance, completion bonus, subtask bonus, and effort) alongside the total accumulated reward (Figure~\ref{fig:reward_metrics}). This decomposition is critical for closing the design loop: it allows researchers to see if an agent is ignoring the task to minimize effort (effort dominance) or moving erratically to maximize speed (distance dominance), facilitating informed tuning of the weights exposed in the GUI (Section~\ref{sec:GUI}).

By default, both performance and reward metrics are directly calculated from the training rollouts used to update the policy, i.e., they display the average value among all episodes generated since the last policy update step. This enables an efficient and direct inference of the current training performance, without additional evaluation overhead. %
To enable a more rigorous, comparable, and reproducible assessment of an (intermediate) policy's performance and quality, we additionally enable adding regular \textbf{evaluations} at a user-defined frequency (e.g., every 2M steps) during the training. %

Finally, for qualitative assessment of learned behavior, we include two video renderings of the trained policy from different camera perspectives in the \textbf{final policy} section. 
Videos of intermediate training policies can also be generated from evaluation runs. These qualitative samples provide the necessary context to interpret the quantitative data, ensuring the learned behaviors are not only successful, but also biomechanically plausible.
Since this comes at the expense of increased training time, it is disabled by default.

\subsection{Example Scenarios}\label{sec:example-demonstrations}
To demonstrate the expressivity of \textit{MyoInteract’s} compositional primitives (DP2), we implemented three diverse HCI scenarios: \textit{AR Interaction}, vertical \textit{Public Display Interaction}, and horizontal \textit{Mobile Keyboard Typing}. We also provide a walk-through on how the GUI can be used to create the \textit{AR Interaction} task. The exact implementation details are provided in the appendix.

\subsubsection{AR Interaction}
\begin{figure*}
    \centering
    \subfloat[\centering First Person View]{{\includegraphics[width=7cm]{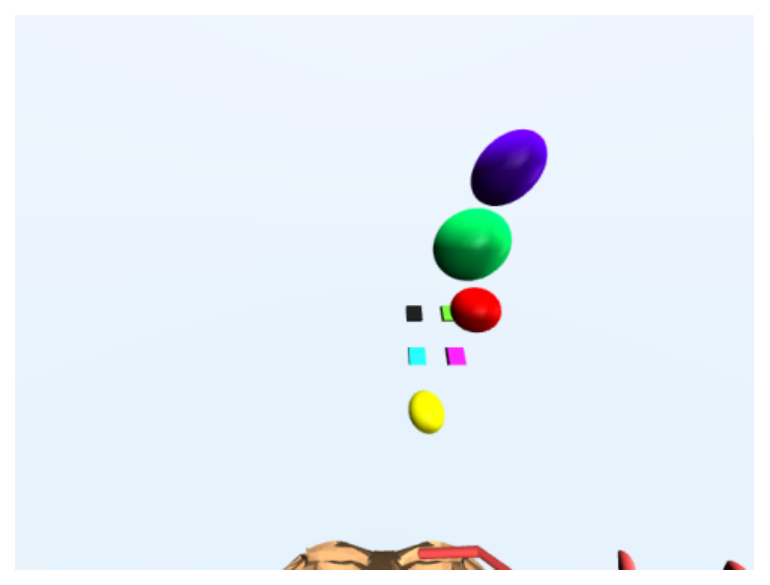} }}
    \qquad
    \subfloat[\centering Side View]{{\includegraphics[width=7cm]{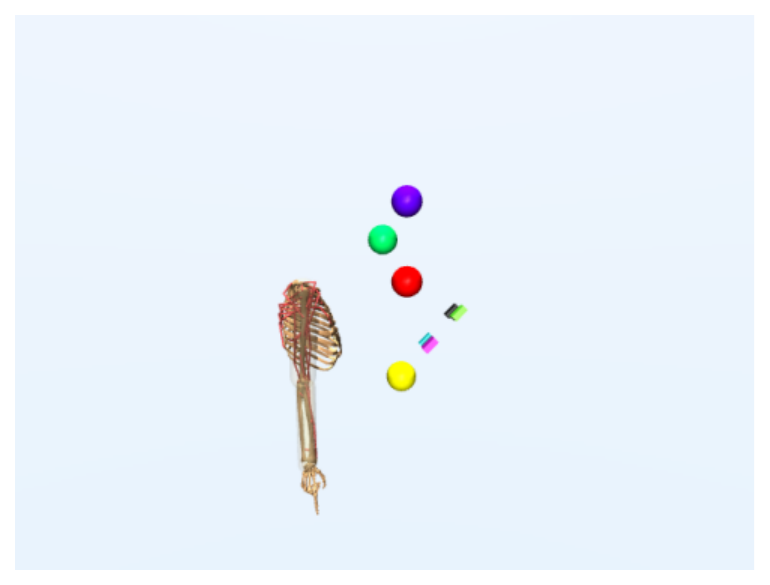}}}
    \caption{Example illustration of an augmented-reality (AR) task where the user has to sequentially complete pointing and button clicking tasks.}     \label{fig:ar_task}
    \Description{An illustration of an augmented-reality (AR) task, presented in two views: a first-person view and a side view. The first-person view, labeled "(a)", shows a user's perspective with the skeleton at the bottom, facing a set of colorful targets, including spheres and small boxes. The side view, labeled "(b)", provides a different perspective, showing a human-like skeleton hand hanging down, also facing the same set of colorful targets. The targets appear to be floating in mid-air, and the user is required to sequentially complete pointing and button-clicking tasks to interact with them.}
\end{figure*}
Consider a designer prototyping an Augmented Reality interface where users alternate between virtual UI elements and physical controls---a common pattern in mixed reality workstations. How does interleaving physical and virtual targets affect movement strategies? Figure \ref{fig:ar_task} shows  an example \textit{AR task} constructed in \textit{MyoInteract}, consisting of four virtual pointing targets and four physical buttons. The following is a quick walk-through of how users would use the GUI to create this task and train an RL agent:

Using the Simple Mode, the user decides the number and types of targets to be 8 (4 physical buttons and 4 virtual spheres),
 along with task-relevant parameters such as button positions and spawn regions for virtual spheres (step 2 in Figure~\ref{fig:teaser}).
Before training, the environment is previewed to confirm that the sub-task targets are situated in reasonable locations, i.e., not beyond the limits of the arm or in difficult-to-reach positions such as behind the user. 
Reward weights can then be adjusted, for example, to emphasize effort minimization or task speed (step 3 in Figure~\ref{fig:teaser}). %
Training is launched (step 4 in Figure~\ref{fig:teaser}) and monitored in real-time using sub-task metrics and reward component plots (step 5 in Figure~\ref{fig:teaser}). After 2 million steps, these logs show (see Figure~\ref{fig:task_progression_metrics}) that the first two targets are reliably reached, while learning is still ongoing for the third, indicating that the current configuration is suitable and training can continue. 
Conversely, flat reward curves early in training signal misconfigured tasks or rewards, prompting early termination and iteration. After training, automatically generated videos of the learned behavior support qualitative assessment of the resulting movement strategies (step 6 in Figure~\ref{fig:teaser}).

\subsubsection{Public Display Interaction}
Public kiosks, ATMs, and information displays require interaction with vertical touch surfaces, raising questions about how vertical target placement affects reachability and movement effort. Figure~\ref{fig:four_buttons} illustrates an example public display task modeled with \textit{MyoInteract}. The simulated user is tasked with pressing the buttons in the pre-specified order. %
By varying the height and spacing of the targets in the GUI, this setup allows researchers to rapidly estimate effort costs for different kiosk layouts without building physical mockups.

\subsubsection{Mobile Keyboard Typing} 
\begin{figure}[!ht]
    \centering
    \subfloat[\centering In the public display task, the simulated user has to sequentially hit 4 buttons on a hypothetical public display. The grey display board is only shown for visualization and does not affect the learned policy. \label{fig:four_buttons}]{{\includegraphics[width=7cm]{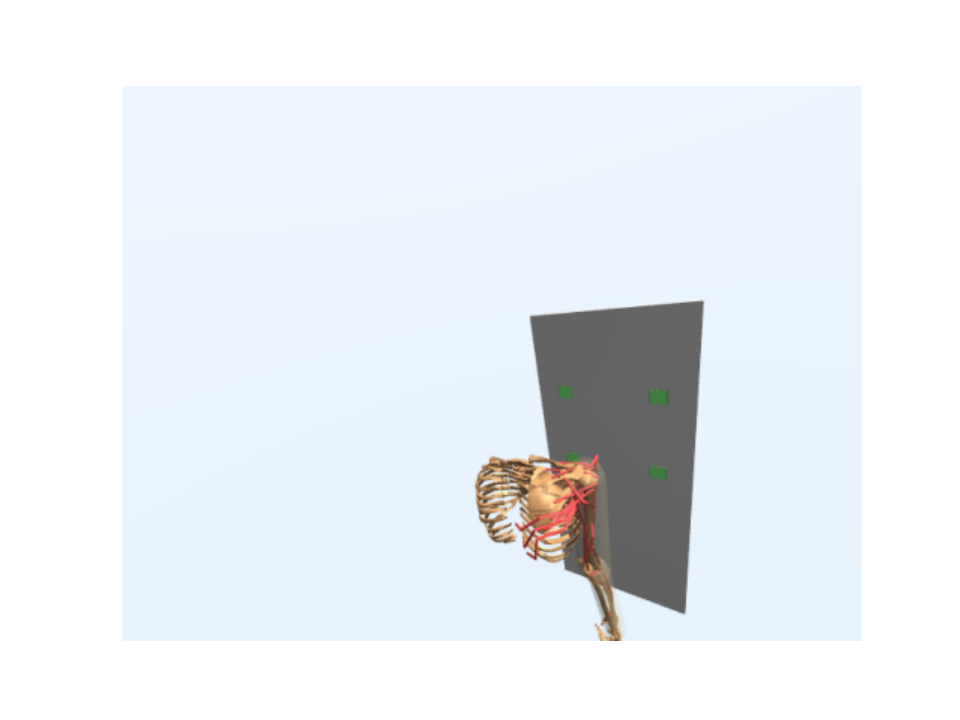}} }
    \qquad
    \subfloat[\centering In the mobile typing task, the simulated user has to sequentially point to each of the five randomly spawned targets on a virtual touch-screen surface resembling the size of an iPhone~17. \label{fig:mobl_typing_image}]{{\includegraphics[width=7cm]{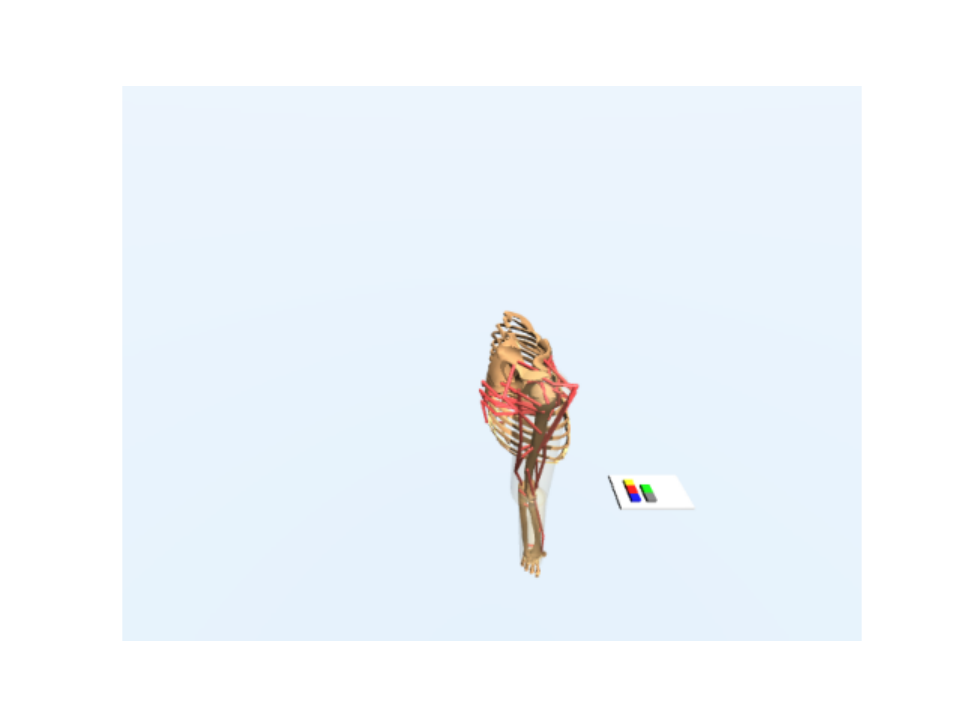}}}
    \caption{Example illustrations of the public display interaction and mobile keyboard typing tasks generated with \textit{MyoInteract}.}     \label{fig:mobl_typing_phones_diagram}
    \Description{Two illustrations of interaction tasks generated with MyoInteract. The first illustration, labelled "(a)", shows a public display task where a skeleton faces a grey vertical display with four green buttons on it. The second illustration, labelled "(b)", depicts a mobile typing task where the skeleton is in front of a white horizontal display resembling the size of an iPhone~17 with five colored buttons on it.}
\end{figure}

Text entry on smartphones involves rapid, sequential pointing to small targets on a horizontal surface held below shoulder height. Figure~\ref{fig:mobl_typing_image} illustrates how \textit{MyoInteract}’s combinatorial task setup can chain multiple pointing targets on a planar surface to simulate touchscreen typing. Target positions are sampled from user-defined ranges; by setting identical upper and lower bounds for the vertical axis, we constrain all targets to a single horizontal plane matching the screen surface.
The depicted phone surface and screen elements (resembling an iPhone~17) are illustrative only, while the interaction logic is handled entirely by the standard pointing primitives.

\section{System Verification}\label{sec:demonstrations}
In the following, %
we demonstrate the speed improvement that \textit{MyoInteract} provides over the current state-of-the-art in biomechanical interaction task simulations \textit{User-in-the-Box (UitB)}~\cite{ikkala_breathing_2022}, and verify that \textit{MyoInteract} predicts biomechanically plausible movements.

\subsection{Fast Biomechanical RL Simulations}

To benchmark the training duration of biomechanical user simulations, we use the standard pointing task and adhere to the task configuration as implemented in the UitB codebase\footnote{\url{https://github.com/User-in-the-Box/user-in-the-box}}, i.e., the simulated user is asked to point to 10 targets in a row, without a reset in between. %
All experiments were conducted on identical hardware (NVIDIA RTX 5090 GPU). For each training, we determined the number of steps until the policy converged based on manual inspection of the reward logs. We additionally ensured that the resulting policy achieved a success rate of at least 95\%, which was the case for all runs.

Table \ref{table:pointing_speed_comparisons} summarizes the results.
On our hardware, training the UitB pointing task took 6.9 hours on average---already faster than the originally reported 24--48 hours in~\cite{ikkala_breathing_2022}, likely due to the use of more recent GPUs. 
To assess the upper bound of speed achievable within our system, we re-implemented the pointing task from \cite{ikkala_breathing_2022} as closely as possible within our framework, ensuring that the reset methods, reward function, and control type closely match those of the original work. The only notable differences arise from slightly different implementations of the PPO algorithm (MyoInteract uses the Brax~\cite{brax2021github} RL library, while UitB relies on Stable Baselines 3~\cite{stable-baselines3}), different RL hyperparameters due to increased parallelization, and the direct provision of information about the target state instead of visual inference.
This configuration, which we call \textit{MyoInteract (Replication)}, \textbf{completes training in ten minutes on average}---a 98\% reduction in training duration (more than 40x faster than the original). %

However, this tuned environment is highly specific to the pointing task and does not generalize well to other types of interaction. To support broader use cases, we also evaluated our \textit{MyoInteract (Default)} configuration, which employs the combinatorial task setup and reward formulation described in Section~\ref{sec:our-methods}. This configuration supports a wider range of sequential, movement-based interactions---such as pointing combined with button clicking---without task-specific tuning.
Even with this more general and thus more “usable” setup, training completes in approximately 36 minutes, representing a 91\% reduction in training time (more than 10x faster) compared to UitB’s 6.9-hour baseline.

\begin{table*}[!ht]
\begin{tabular}{@{}lccccc@{}}
\toprule
Experiment             & Num Runs & Num Steps (M) & Total Training Time (hrs)      \\ \midrule
UitB (On Older Hardware) ~\cite{ikkala_breathing_2022}                     & -        & 40--80           & 24--48                    \\
\midrule
UitB (On Our Machine)                            &    3      &   42.9 $\pm$ 9.0            & 6.9 $\pm$ 1.5                     \\

MyoInteract (Replication)                          & 5          &  7.4 $\pm$ 0.8             &  0.17 $\pm$ 0.03                   \\ 
MyoInteract (Default)            & 5         &   34.9 $\pm$ 1.9            & 0.60 $\pm$ 0.03                      \\

\bottomrule
\end{tabular}
\caption{A comparison of training steps and wall times required for training until convergence
in the default pointing task~\cite{ikkala_breathing_2022} (mean $\pm$ std. observed across the respective number of runs).}
\label{table:pointing_speed_comparisons}
\Description{Comparison of training efficiency for a pointing task across four settings. UitB on older hardware required 40–80 million steps and 24–48 hours. UitB on our machine took 42.9 ± 9.0 million steps and 6.9 ± 1.5 hours over 3 runs. MyoInteract replication completed in 7.4 ± 0.8 million steps and 0.17 ± 0.03 hours over 5 runs. MyoInteract with default settings required 34.9 ± 1.9 million steps and 0.60 ± 0.03 hours over 5 runs.}
\end{table*}
In Table \ref{table:example_cases}, we also show that the three demonstration examples introduced in Section \ref{sec:example-demonstrations} can successfully be trained in less than 30 minutes.

\begin{table*}[ht]
\begin{tabular}{@{}lccccc@{}}
\toprule
Experiment     & Num Subtasks & Num Runs & Num Steps (M) & Total Training Time (min)     \\ \midrule
AR             & 8           & 5         &      9.9 $\pm$ 1.3     &    23.2 $\pm$ 2.2  \\
Public Display & 4           &   5       &     6.7 $\pm$ 2.5      & 14.9 $\pm$ 4.9                      \\ 
Mobile Typing  & 5           &  5        & 10.4 $\pm$ 4.1          &   11.7 $\pm$ 3.9                  \\
\bottomrule
\end{tabular}
\caption{Training time and steps required to train until convergence
for three example interaction scenarios (mean $\pm$ std. observed across the respective number of runs).}
\label{table:example_cases}
\Description{Training results for three interaction scenarios. AR used 8 sub-tasks over 5 runs, requiring 9.9 ± 1.3 million steps and 23.2 ± 2.2 minutes. Public Display used 4 sub-tasks over 5 runs, requiring 6.7 ± 2.5 million steps and 14.9 ± 4.9 minutes. Mobile Typing used 5 sub-tasks over 5 runs, requiring 10.4 ± 4.1 million steps and 11.7 ± 3.9 minutes.}
\end{table*}
\subsection{Motion Evaluation}

\begin{figure*}
    \centering
    \subfloat[\centering Fitts' Law Evaluation \label{fig:fitts_law}]{\includegraphics[width=0.3\linewidth]{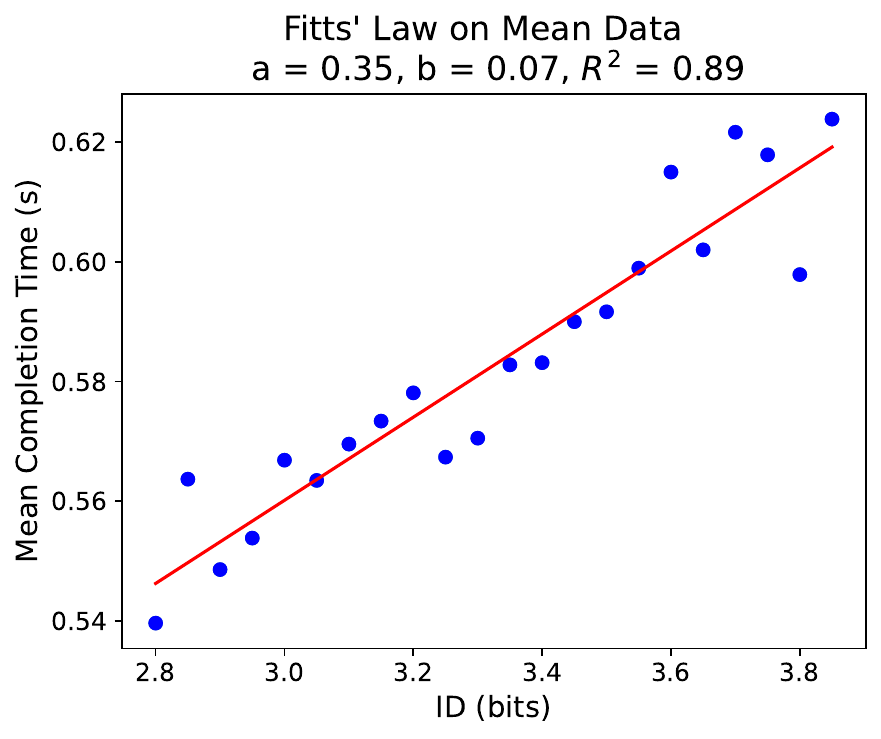}}
    \qquad
    \subfloat[\centering Projected Positions \label{fig:projected_positions}]{{\includegraphics[width=0.3\linewidth]{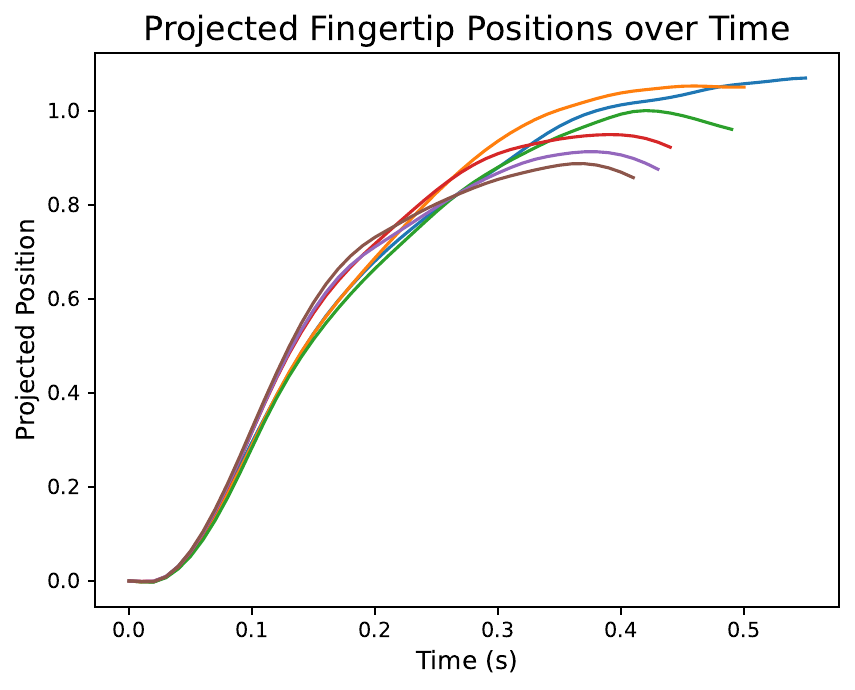} }}
    \qquad
    \subfloat[\centering Projected Velocities \label{fig:projected_velocities}]{{\includegraphics[width=0.3\linewidth]{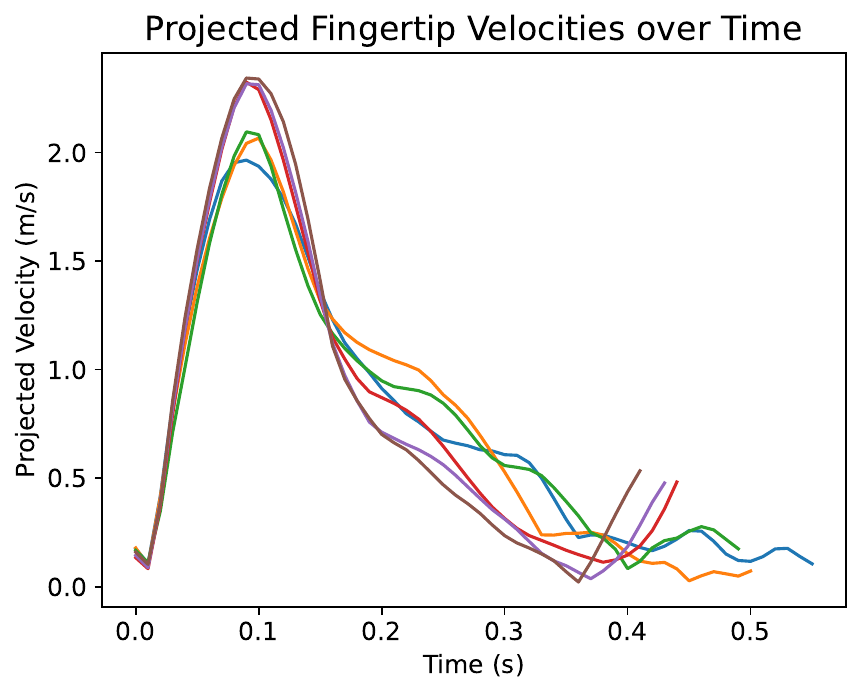} }}
    \caption{
    Biomechanical validation of the policy trained using \textbf{MyoInteract (Default)}.\\
    (a) Average movement times (blue dots) plotted against IDs of targets. The observed movement times follow a clear linear relationship (red line), as predicted by Fitts' Law, with a fit of $R^2 = 0.89$.
    (b) Projected fingertip positions for a trained policy, moving from the initial position (0) to the target position (1) and settling there, 
    with the different lines corresponding to six different target widths.
    (c) Corresponding projected fingertip velocities, exhibiting the characteristic bell-shaped profile of aimed reaching movements~\cite{morasso1981spatial}.}
    \label{fig:projected_pos_vel}
    \Description{Validation of a trained policy using biomechanical evaluation, consisting of three plots. The first plot, labeled ``Fitts' Law Evaluation'', shows a scatter plot of average movement times against target IDs, with a linear line illustrating fitted model predictions (red line) and a fit of R2=0.89, indicating that the observed movement times follow Fitts' Law.
    The second plot, "Projected Fingertip Positions," displays the projected fingertip positions over time, illustrating a smooth movement from the initial position (0) to the target position (1) and stabilizing at the goal.
    The third plot, "Projected Fingertip Velocities," shows the corresponding projected fingertip velocities, exhibiting a characteristic bell-shaped profile of aimed reaching movements.}
\end{figure*}

We also examine whether the movement policies generated by MyoInteract follow established movement regularities such as Fitts' Law. We use the \textit{MyoInteract (Default)} policy trained on the default pointing task and evaluate it across a range of indices of difficulty (IDs), using a sampling frequency of 100 Hz~\cite{ikkala_breathing_2022}. 
As shown in Figure~\ref{fig:fitts_law}, the policy trained with our framework adheres to Fitts' Law ($R^2=0.89$).
Moreover,
the fingertip moves smoothly toward the target and remains close for the set dwell time (Figure \ref{fig:projected_positions}), and %
the velocity profiles exhibit the typical bell-shaped pattern~\cite{morasso1981spatial}, followed by a second, smaller corrective submovement required to keep the fingertip inside the target for the required dwell duration (``settling phase'') (Figure \ref{fig:projected_velocities}). %
These results are consistent with previous findings~\cite{ikkala_breathing_2022, fischer2021reinforcement, moon_real-time_2024}, demonstrating a comparable level of movement regularity to previous biomechanical simulations, but at much faster speeds. %
Similar results were obtained for the \textit{MyoInteract (Replication)} policy (see Appendix \ref{sec:validation_replication}).

\

\section{Workshop Evaluation}\label{sec:workshop-evaluation}
To assess the usability and potential of our framework and GUI, we conducted an evaluation study with 12 HCI researchers, most of them with little to no prior knowledge in RL and computational user simulations. 
We were mainly interested in whether participants were able to setup and run trainings, and how useful the provided GUI features and metrics were in monitoring and assessing the quality of learned user behavior. In addition, we asked participants about their opinion of biomechanical RL as an HCI research and design method, and how they would integrate biomechanical user simulations into their workflow.

\subsection{Methodology and Procedure}
\label{sec:workshop_eval_method}

The study was exploratory in nature and primarily focused on qualitative findings, complemented by selected quantitative measures. 
It was approved by the departmental ethics committee in advance.
The study was conducted as part of a two-hour workshop session on biomechanical RL, which included a brief tutorial (20 mins) on the basics of biomechanical modeling, reinforcement learning, and reward function design, ensuring participants had a common foundational understanding before engaging with the study tasks.

Each participant completed the study individually following a structured procedure. After the tutorial, each participant received a link to the graphical user interface described above, running on a cloud-hosted server with GPU access (NVIDIA RTX 5090). %
They were then asked to open the GUI and were given a short introduction into the possible task setups and other parameters that can be modified in the simple mode.
Afterwards, each participant was asked to define an interaction task of their choice, including up to 10 subtasks. 
Participants with a specific interaction or interface concept in mind were encouraged to recreate it using the provided tools. Participants were allowed to use one of the preloaded configurations (see Section~\ref{sec:example-demonstrations}), but then had to modify at least two task parameters. 
While the simple GUI was sufficient for completing the task, more experienced users could optionally explore the advanced parameter settings (see Section~\ref{sec:GUI}).

Participants were given 15 minutes to configure their desired task, adjust relevant parameters, and start training. Afterwards, they were instructed to verify that the training was running correctly and that the results were being logged as expected. During training, which lasted between 20 and 60 minutes for their chosen configurations, they could either take a break or monitor progress using the provided metrics (Section~\ref{sec:metrics}). After training was completed, participants spent approximately 5 minutes evaluating the resulting policy based on the generated plots and videos (steps 5 and 6 in Figure~\ref{fig:teaser}) and reflecting on how the policy might be further improved. %

Following the practical task, participants completed an online questionnaire consisting of open-ended questions and 5-point Likert scales. The questionnaire included both standardized instruments (demographic questions)
and task-specific items addressing (1) the process of defining and training RL tasks, (2) the usability of the interface for policy setup and training, and (3) the monitoring and evaluation of trained policies using WandB. Participants were also asked to reflect on whether the trained policy met their expectations and how they might adjust parameters in a subsequent iteration. An overview of all questions can be found in Appendix~\ref{sec:study-questions} and the anonymized responses are included in the auxiliary material.

In the final stage of the study, participants took part in a group discussion on the potential use cases of the framework, its relevance to their own research or projects, and desired features for improving its usefulness.
The group discussion was recorded with participants' consent and subsequently transcribed. 

\subsection{Participants}
This workshop study included 12 participants (9 male, 3 female) with an average age of 28.5 years. Their levels of expertise ranged from novice to expert across reinforcement learning (8 novices, 3 intermediate, 1 expert), biomechanical modeling (9 novices, 2 intermediate, 1 expert), and cognitive modeling (4 novices, 7 intermediate, 1 expert). Participants were not compensated for their participation.

\subsection{Results}
We structure our results across the themes we identified from participants' responses. 

\paragraph{Reinforcement Learning Setup and Training}

Participants generally found the interface intuitive and accessible for both defining a reinforcement learning task and training a simulated user. 
Participants were able to configure and manipulate tasks effectively, fostering an accessible and engaging learning experience: \textit{``The interface made it easy to input necessary parameters without requiring extensive background knowledge in RL. The parameter setup was intuitive and accessible, allowing me to configure the task even as a non-expert in this field.''} (P12). Participants also appreciated the system’s guided workflow: \textit{``I found it useful that the interface guided me step by step and presented the process in a straightforward way. The clear sequence of actions and visual feedback made the training easy to follow and reduced confusion during the setup.''} (P5). 

Several participants highlighted certain GUI elements. For example, P11 mentioned, \textit{``defining a task was intuitive, especially because there was immediate visual feedback available in the notebook cell''}, while P5 found that, \textit{``this visual interaction helped me understand how different parameter settings might influence the task design''}. %
The use of parameter sliders was consistently considered intuitive and informative, enabling participants to explore meaningful value ranges without prior knowledge (P3, P5, P7). For example, P5 emphasized that sliders \textit{``made it very clear and intuitive to define the Reinforcement Learning task, since I could easily adjust values without needing to type or remember specific commands''}. Users also found the customizable configuration of object properties (P9) and the comprehensive parameter view (P10) particularly useful, supporting an efficient task setup.
Additional features such as reward function visualization %
and parameter configuration via dials were seen as valuable tools for refining training tasks (P8, P11).

Quantitatively, 
participants rated the process of defining and setting up a reinforcement learning task as relatively easy (mean = 3.58, SD = 0.90) and not particularly difficult (mean = 2.33, SD = 0.89). Likewise, training a simulated user was perceived as easy (mean = 3.75, SD = 0.97) and not difficult (mean = 2.18, SD = 0.75). These ratings indicate that participants were generally able to engage effectively with the system and complete the setup and training without major obstacles.

\paragraph{Policy Monitoring and Analysis}

Participants emphasized the need for clear and interpretable \textbf{visualizations} to effectively monitor training progress and assess policy quality. Video demonstrations and visual footage of the simulated user's behavior were consistently regarded as intuitive means of understanding policy performance (P3, P10, P11, P12). For example, P11 found, \textit{``the visual footage helpful to intuitively get a sense of how successful the policy was''}. 
Participants relied on success rates, loss curves, episode lengths, and reward component values to track learning progress and evaluate training outcomes (P1, P5, P6, P7, P8).

While some participants noted that the existing metrics were already sufficient for their analysis needs (P7, P8), others suggested extensions %
to further improve monitoring and analysis. %
These included visualizing intermediate policies during training to illustrate performance improvement over time (P1), %
incorporating additional metrics such as arm movement speed (P4), %
and providing clearer descriptions and prioritization of metrics to help users focus on the most relevant information:
\textit{``In general I would have needed way less metrics and a focus on the most important ones would have been nice. Also a better description of the metrics, so what exactly am I seeing here.''} (P8). Other suggestions included reward \textit{space} visualization using a Principal Component Analysis (P10) and methods to link metrics directly to observed agent behavior for more intuitive interpretation (P11). 

Quantitative ratings reflected these qualitative findings. Participants found it slightly easier on average to assess the quality of the trained policy (mean = 3.75, SD = 1.14) than to monitor the training progress (mean = 3.42, SD = 1.08), with corresponding difficulty ratings of 2.25 (SD = 1.22; assessing) and 2.67 (SD = 0.99; monitoring), respectively.

\paragraph{Policy Performance and Next Steps}
Most participants reported being satisfied with the performance of the trained policy (P1, P2, P5, P7, P9, P10, P11, P12). Several participants proposed refinements to further improve results. Most suggestions focused on modifying the reward function weights, e.g., to reduce motion jitter or encourage faster, more natural movements (P1, P4, P7). %
Others emphasized changes in the task configuration, such as adjusting the target distribution (P10), using \textit{``more uniform target settings for each targets to make some generalized human motion for diverse ranges''} (P10), \textit{``narrowing down the sampling coordinates a bit more''} (P4), or reducing task complexity (P8). 
Given that the policies already achieved strong results, some participants expressed interest in further exploring the boundaries of the simulated user: \textit{``I hope to try out more bizarre/ambitious training setup next time.''} (P11), for instance by using smaller objects or targets placed at larger distances (P3). One participant noted that \textit{``It was beyond my expectation. It seemed to have understand physical constraints well because sometimes movement was blocked by torso.”} (P9).

\paragraph{Suggestions for UI Improvement}
Participants identified several opportunities to improve the interface and further simplify both the definition of tasks and the training of simulated users.
A key theme concerned the need for more intuitive and informative visual feedback (P1, P3, P7, P8, P9, P10, P11). Participants suggested aligning parameter sliders with a live scene preview featuring automatic rendering: \textit{``I would rather have a direct manipulation interface for adjusting object positions, or at the very least put the object sliders side-by-side with the scene preview and have it auto-render.''} (P1), or showing 3D views: \textit{``It would be better if I could set the parameters with corresponding 3D GUI, which can make the usage more intuitive.''} (P10). 

Participants also emphasized that additional guidance during task configuration, such as more recommendations on useful parameter ranges or default values, could help further reduce uncertainty and support informed decision-making (P5). %
One participant suggested adding a preview function to completely eliminate the need for full training in certain cases: \textit{``visualization of expected changes of how the change in parameters would influence the learning task and training would help understand.''} (P3).
Other participants requested improved real-time training feedback (P11, P12) and the ability to adjust parameters \textit{during} training (P4).
Suggestions include %
adding interface elements such as a progress bar or estimated completion time, as well as clear indicators showing whether the training is running or visualizations are being generated (P12).
Participants also recommended prioritizing a small set of core metrics (e.g., loss, accuracy, precision) to support clarity and reduce cognitive load, particularly for non-expert users.

\paragraph{Potential Use Cases}
Participants discussed the practical potential of MyoInteract. One participant observed that \textit{``in the real world of medical service or games or whatever, there are some actual behaviors that is resembled to those simple button-pressing behaviors''}, while another highlighted the PlayStation joystick as an example of a forearm-bending task comparable to those studied. These comments motivated suggestions to embed additional real-world tasks and input devices into the framework.

\section{Discussion}\label{sec:discussion}

Previously, training simulated users via biomechanical RL took hours to days and required substantial expertise in computational user models and how they are implemented. Researchers who wanted to test a design hypothesis or low-fidelity prototype in silico could not resort to a default user model and train it ``out of the box'', but had to first familiarize themselves with the implementation and code details of complex simulation frameworks and physics engines such as UitB~\cite{ikkala_breathing_2022}, MyoSuite~\cite{caggiano2022myosuite}, or MuJoCo~\cite{todorov_mujoco_12}, before they could setup and start hour-long training runs that lack substantial feedback on the progress and quality of the simulation. 
Our main motivation for developing MyoInteract was to bridge these \textit{gulfs of execution and evaluation}~\cite{Norman2002-da} and make biomechanical RL simulations accessible to researchers in the field of interaction design. In the following, we reflect on the extent to which MyoInteract fulfills these design goals, the potential it unleashes for HCI research and interaction design prototyping, and lessons learned that we believe generalize to biomechanical user simulations in the scope of HCI. We conclude with a discussion on current limitations and future extensions.

\subsection{Reflections on the Design Goals} %
We designed \textit{MyoInteract} to lower the entry barrier to biomechanical RL for interaction design research by tackling three design goals (DGs). Using Norman's action cycle~\cite{Norman2002-da}, this involves addressing the gulfs of execution (DG1) and evaluation (DG2) while also reducing the temporal costs that compound both gulfs (DG3). %
The workshop study provides evidence of how well the system met these three design goals and reveals important limitations and opportunities for improvement.

A primary bottleneck in prior biomechanical RL workflows is the time required to train a policy for a given interaction task, which often forces researchers to %
wait hours or days before assessing its predictions. Consequently, we %
built upon recent advances in GPU-accelerated physics simulation~\cite{thibault2024learning, zakka2025mujoco} and developed a biomechanical framework that achieves substantial training speed-ups through massive parallelization.
In fact, our implementation of the standard pointing task from~\cite{ikkala_breathing_2022} reduced the
\textbf{training time (\ref{dg-speed})} by a factor of 10--40x. Moreover, all workshop participants were able to complete their training runs within 20--60 minutes, across a range of task configurations. %
Notably, the improvements in speed did not compromise the biomechanical plausibility of the predicted user trajectories, as was demonstrated in Section~\ref{sec:demonstrations} through a Fitts' Law study and a qualitative analysis of the predicted target acquisition and selection movements.
This shows that MyoInteract has successfully achieved design goal~\ref{dg-speed}. %

Moreover, we applied the design principles of \textit{Abstraction via Task Decomposition}, \textit{Visibility via GUI}, \textit{Constrained Interaction}, and \textit{Progressive Disclosure} (see Section~\ref{sec:design_principles}) to reduce the \textbf{gulf of execution (\ref{dg-execution})}. %
Workshop participants described the GUI as intuitive and accessible, and several participants remarked that they were able to configure and run trainings ``even as a non-expert in the field''. Features particularly appreciated include GUI sliders (DP3), (adaptive) default values (DP3, DP5), and the guiding structure and modes of the GUI (DP6), which allowed the participants to focus on testing different interactions and task configurations that interested them, rather than getting bogged down in technical implementation details.
Feedback from participants also revealed current limitations of MyoInteract.
While the GUI provides renderings of the environment from two different camera perspectives, several participants expressed an interest in an interactive 3D view of the interaction environment that enables direct manipulation of target positions and sizes.
Similarly, being able to setup target configurations interactively via sliders, as provided by our GUI, is an advantage over using Python scripts and config files; however, our study revealed a clear demand for more direct and ``natural'' manipulation methods. In summary, MyoInteract was successful in reducing the gulf of execution, while the workshop study revealed promising directions for further improvements that need to be investigated in more detail.

To reduce the \textbf{gulf of evaluation (\ref{dg-evaluation})}, MyoInteract provides \textit{Multi-Level Feedback} (DP4) before, during, and after training, including environment previews, decomposed metrics, and videos showing the learned behavior.
Participants placed a high value on the provided visual feedback. In particular, videos of trained policies were considered crucial for evaluating the performance of trained policies. 
Decomposed metrics, such as per-target success rates and individual reward components, enabled analytical inspection and allowed participants to identify potential shortcomings and propose parameter improvements for future iterations.
From the workshop study, we learned that intermediate visualizations during the training can complement these metrics, resulting in a comprehensive set of options for monitoring and assessing simulation results. %
Moreover, some participants found the number of available metrics overwhelming and expressed a desire for stronger guidance on how to interpret these quantities. %
These findings suggest that our approach effectively reduced the gulf of evaluation by exposing relevant training feedback, but that further addressing this gulf also requires careful consideration of when and how such feedback is presented, which should be further examined.

Overall, the workshop evaluation indicates that MyoInteract is making meaningful progress toward increasing the usability and utility of biomechanical RL for interaction design, while highlighting specific areas that can be further improved.  %
The feedback from participants provides valuable guidance for refining the balance between speed, abstraction, and interpretability in future iterations of the framework. 

While the workshop demonstrated that novices can learn and operate the framework within a single session, evaluating the framework's long-term impact on real-world prototyping of interactive systems remains an open question for future work.

\subsection{MyoInteract as an Enabling System for HCI Research and Beyond}

Biomechanical RL's potential for HCI has been demonstrated in isolated studies, yet its broader value remains unproven---largely because the barriers to entry have prevented widespread exploration. 
By removing these barriers, %
our framework lowers both the expertise and computational thresholds for entry, and makes biomechanical RL an interactive tool for HCI researchers to test a variety of interaction designs in silico. 
For example, MyoInteract can be used to investigate the ergonomics of different UI layouts within an XR application, learn about the effect of design choices on users' movements (e.g., to what extent they bend the elbow or move the arm back and forth), or identify the optimal design of a set of buttons (location, size, physical properties), ensuring reachability, precision, and performance.
More generally, we believe that MyoInteract can be useful for testing any interaction designs that involve an infinite parameter space and where ergonomics and biomechanical measures, such as movement patterns or physiological effort, are essential design factors. 

Moreover, MyoInteract can be used to assess and improve the quality of biomechanical user simulations. %
Given that our framework offers a large range of options for adjusting parameters of the simulated user and provides a fast and usable (though clearly simplified) workflow for biomechanical RL simulations that enhances the current state of the art, we believe that MyoInteract will be valuable not only to the HCI community, but also to simulation engineers and researchers developing computational models of user behavior. Ultimately, MyoInteract has the potential to catalyze collaborations between interaction designers and other disciplines adjacent to HCI, including computational user modeling, cognitive psychology, sports and rehabilitation, and neurorobotics.

\subsection{Design Reflections for Biomechanical User Simulations for HCI Tasks} %
The design and evaluation of MyoInteract revealed insights that transcend this framework and extend to biomechanical user simulations in the broader context of interaction design and prototyping.

\textbf{Iteration costs shape feasible opportunities}. Reducing training time from days to minutes did not merely accelerate existing workflows---it enabled qualitatively different ones. %
Rather than committing to a single configuration, %
users can run multiple simulations within a single session, using outcomes from one run to inform the next.
This shift enables exploratory behaviors such as comparing task variants, probing edge cases, and incrementally refining rewards or layouts---activities central to interaction design but previously infeasible with long training cycles.
More generally, in biomechanical user simulations, the cost of iteration determines not only the efficiency, but also whether iterative exploration is possible at all.

\textbf{Lowering interaction friction accelerates learning without removing complexity.}
MyoInteract demonstrates that lowering entry barriers is less about simplifying underlying models than about how they are exposed. Making options visible, defaults explicit, and ranges constrained enables productive use before full conceptual understanding. Progressive disclosure supported task setup, but participants’ responses also showed that unprioritized metric visibility can be overwhelming. This suggests that reducing friction in biomechanical user simulation frameworks requires not only accessible interfaces, but deliberate curation of what information is surfaced, and at which stage of the workflow.

\textbf{Black-box methods demand intermediate and immediate observability.} Biomechanical simulation methods remain difficult to interpret, especially when training fails or produces unexpected behavior. By exposing decomposed metrics during training, MyoInteract supported systematic diagnosis rather than ad-hoc speculation, and enabled earlier intervention when runs were clearly unproductive. Simulation methods that operate as black boxes should provide intermediate, interpretable feedback to support sense-making and debugging during training, rather than deferring insight to post-hoc evaluation.

\textbf{Expressivity and usability trade off against each other.} Configuration-driven task authoring allowed non-programmers to create simulations, but necessarily constrained the space of expressible interactions.
Similar tensions surfaced throughout the system: exposing more parameters or metrics increases potential insight and control, yet raises cognitive load and can hinder effective use. These trade-offs have no universal solution and require context-dependent design decisions. Our workshop highlights both successful choices and missteps, underscoring the importance of iterative refinement grounded in user feedback.

\subsection{Limitations and Future Work}

Although MyoInteract shows great promise for HCI research and interaction design, there are some clear limitations that future work should address.

\textit{MyoInteract} provides three pre-configured example scenarios with default hyperparameters that produce stable, convergent training. However, certain parameters, such as reward weights, observation configurations, and training duration, will not lead to successful training across all possible combinations. In particular, reward weights require careful tuning: poorly balanced weights can cause training to converge slowly or produce unintended behavior. 
We currently provide an adaptive training step scaling, which scales the training duration based on the number of targets and dwell time. While this is better than setting an arbitrary step count that risks overtraining or undertraining, it does not guarantee convergence for all configurations.
Additionally, the current observation space does not encode the positions of future targets, and if the agent is required to navigate around these targets, the observation space will have to be adapted.
A systematic evaluation of which parameter combinations produce reliable training across a wider range of tasks remains important future work.

Notably, the range of interaction tasks captured by \textit{MyoInteract} is currently limited to sequential target acquisition and selection via direct manipulation. 
While direct and sequential target selection forms the basis of most GUI-based interaction and allows investigating the ergonomics of XR, mobile, and other touch- and clicking-based interfaces, future work should further aim to increase the task scope. Suggestions from our workshop participants include modeling input devices such as joysticks or VR controllers and indirect manipulation methods such as raycasting or Go-Go~\cite{poupyrev1996go}.

In addition, our framework currently only focuses on biomechanical aspects of human-computer interaction. While these are essential for a range of movement-based interaction tasks~\cite{bachynskyi2015performance}, extending musculoskeletal simulations executed in a physics engine such as MuJoCo with computational models of human cognition and perception would enable more holistic simulations that capture the entire sensorimotor control loop~\cite{fischer_optimal_2022}. Future work should explore opportunities for integrating models from computational rationality~\cite{lewis2014computational, oulasvirta_computational_2022, chandramouli_workflow_nodate} with biomechanical RL simulations, and align the observation space with multimodal perception models. This particularly includes modeling how users visually sense and perceive their environment, beyond the prevailing monocular, static, and low-resolution vision model %
currently implemented in simulation frameworks such as UitB~\cite{ikkala_breathing_2022}. %

\textit{MyoInteract} currently trains each configuration from scratch, though our unified observation space and reward structure have the potential to learn skills that transfer across tasks. Future work should investigate this potential, exploring, for example, warm-starts or the training of foundational policies that adapt to unseen tasks without fine-tuning.
Moreover, future work should investigate metrics to detect artifacts such as motion jitter during training and reduce reliance on manual video analysis.

Finally, our framework has only been tested with a single biomechanical model of the upper extremity. 
Notably, \textit{MyoInteract} comes with an experimental hand model based on the \textit{MyoArm} model, which in principle allows creating user simulations for more dexterous tasks such as gamepad control or multi-touch typing.
Moreover, users with more biomechanical expertise can readily add other models, e.g., two-armed upper body models such as the one used in~\cite{Hetzel2021}, or the hand, back, or leg models provided by the MyoSim library\footnote{\url{https://github.com/MyoHub/myo_sim}}, which \textit{MyoInteract} is fully compatible with.
It would be interesting to compare these models in terms of speed, scope, and robustness, and explore how to best integrate these models into \textit{MyoInteract} regarding the design trade-off between complexity and usability.

\section{Conclusion}\label{sec:conclusion}
In this paper, we introduced \textit{MyoInteract}, a novel framework for fast prototyping of biomechanical HCI tasks.
Using the Human Action Cycle as a design lens, we identified three fundamental barriers that affect the accessibility, iterability, and interpretability of biomechanical simulations in HCI. 
We reduce the temporal cost by utilizing GPU acceleration, which reduces training times by up to 98\%, enabling practical iterations for the first time as training times are lowered from days to under an hour, or even just a few minutes. 
We address the gulf of execution by utilizing a combinatorial task setup and GUI to construct a system, which enables exploration of more expressive interaction interfaces than previous systems, while being easy to learn and use.
Multi-level feedback, including real-time metrics, provides a greater degree of insight into the black-box nature of RL training. This enables users to identify and iterate on poorly configured experiments or models early on rather than waiting for flawed runs to terminate, thus tackling the gulf of evaluation.
As demonstrated through a workshop evaluation with 12 participants, \textit{MyoInteract} can be learned by HCI researchers with little to no expertise in biomechanical modeling and RL within two hours, allowing them to successfully setup user simulations for a range of relevant HCI interaction tasks. 
Ultimately, \textit{MyoInteract} does more than just accelerate computation; it accelerates insight. By making the physical body computationally accessible, we hope to establish biomechanics not just as a method for explaining movement-based interaction, but as a generative tool for designing it.

\begin{acks}
    This work was supported by EPSRC grant EP/W02456X/1.
    The authors gratefully acknowledge the computing time made available to them on the high-performance computer at the NHR Center of TU Dresden. %
    Hannah Selder and Arthur Fleig acknowledge the financial support by the Federal Ministry of Research, Technology and Space of Germany and by Sächsische Staatsministerium für Wissenschaft, Kultur und Tourismus in the programme Center of Excellence for AI-research „Center for Scalable Data Analytics and Artificial Intelligence Dresden/Leipzig“, project identification number: ScaDS.AI.
\end{acks}

\section*{GenAI Usage Disclosure}
We have used AI-powered coding tools to help write parts of the codebase used for this research. We have always tested and checked any AI-generated code to ensure that it fulfills the desired functionality. 
\bibliographystyle{ACM-Reference-Format}
\bibliography{sample-base}

\clearpage
\appendix
\section{Technical Details for MyoInteract}
\subsection{Biomechanical Model}
In the biomechanical model adapted from the UitB framework\footnote{\url{https://github.com/User-in-the-Box/user-in-the-box}}, the torso, wrist, and fingers are fixed, resulting in a model with five degrees of freedom---three at the shoulder, one for elbow flexion, and one for forearm pronation-supination---driven by 26 Hill-type muscles. 

Motor execution noise is defined as follows:
\begin{equation} 
 a = \text{clip}_0^1 \left(\left(1 + k_{\text{SDN}} \cdot z \right) \cdot \tilde{a} + k_{\text{CN}}\right), \quad z \sim \mathcal{N}(0,1), 
\end{equation} 
where $\tilde{a}$ and $a$ denote the intended and the applied muscle control signals, respectively, $z$ is a standard normal random variable, and $k_{\text{SDN}}\geq0$ and $k_{\text{CN}}\geq0$ are the selected signal-dependent and constant noise levels. 
 
Muscle controls are updated at a frequency of 20~Hz by querying the current (during training) or trained (after training) policy with the latest observation. For the MuJoCo physics simulation, a higher internal update rate of 500 Hz is used to ensure stability. 

We implement three reset modes for the biomechanical model that determine the initial state at the beginning of each episode: \textit{zero}, \textit{epsilon-uniform}, and \textit{range-uniform}.  In the \textit{zero} reset, all joint angles and velocities are initialized to zero, resulting in a relaxed body pose with the arm hanging down.  The \textit{epsilon-uniform} reset initializes the model with randomly perturbed joint angles and velocities around the zero pose ($\pm 0.05$ rad and $\pm 0.05$ rad/s, respectively).  Finally, the \textit{range-uniform} reset selects the joint angles uniformly randomly from the valid joint range for each joint individually, while the velocities are selected as in the \textit{epsilon-uniform} reset mode.  For all modes, initial muscle activations are randomly sampled within the unit interval. The non-zero resets are intended to reduce overfitting of the trained movements to a specific pose.

\subsection{Task Setup}
We define each pointing target using the following specification:
\begin{itemize}
    \item Size: Each target is a sphere of a randomly-sampled radius $r \in [r_{min}, r_{max}]$, where the user can specify the lower and upper limits $r_{min}$ and $r_{max}$.
    \item Location: Each target location is randomly drawn from a 3D space with boundaries  $[x_{min}, x_{max}] \times [y_{min}, y_{max}] \times [z_{min}, z_{max}]$ specified by the user. %
    \item Dwell duration: The minimum time (in seconds) the end-effector must be kept inside the target sphere to successfully complete the task. By default, this is set to 250 ms (corresponding to 5 time steps).
    \item Color: The color of the target sphere. While the target color does not affect the learned behavior of simulated users without vision, this property can be useful to setup task environments with multiple targets (see above in  Section \ref{sec:example-demonstrations}).
\end{itemize}

We model buttons as MuJoCo box geometries (i.e., cuboids) with the following properties:
\begin{itemize}
    \item Size: Each button is a  cuboid of user-specified size $w \times h \times d$ (width $\times$ height $\times$ depth), where $w$ and $h$ denote the width and height of the button surface with $d$ as its depth.
    \item Location: The 3D location of the center of the button $(x, y, z)$.
    \item Minimum touch force: The minimum force the end-effector must exert on the button to successfully complete the button clicking task.
    \item Orientation angle: The orientation angle of the button around the horizontal $y$-axis (in degrees), defining its tilt.
    \item Color: The color of the button.
\end{itemize}
\subsection{Demonstration Tasks Implementation Details}
\subsubsection{AR Interaction}
We have eight targets consisting of four buttons and four pointing targets, where the buttons are fixed in position while the pointing targets are randomly spawned in a target area. The task consists of repeated pointing and button pressing tasks where the buttons are represented as square boxes of 5 cm face widths with their $[y,z]$ coordinates in meters set to $[0.15, -0.29]$, $[0.15, 0.01]$, $[-0.15, -0.29]$, $[-0.15, 0.01]$, on a flat vertical plane 42 cm in front of the user's shoulder, respectively. The boxes representing the buttons are at 45 degrees around the horizontal and require a minimum touch force of 1 N. The spherical pointing targets, which all have a dwell duration of 250 ms, have their coordinates and radii sampled according to:
\begin{itemize}
    \item Target 1: $x \sim U(0.225, 0.35)$, $y \sim U(-0.15,0.15)$, 
    
    $z \sim U(-0.3, 0.3)$, $r \sim U(0.04, 0.08)$
    \item Target 3: $x=0.3$, $y \sim U(-0.15,0.15)$, 
    
    $z \sim U(0, 0.3)$, $r \sim U(0.04, 0.08)$
    \item Target 5: $x=0.3$, $y \sim U(-0.1,0.1)$, 
    
    $z \sim U(0, 0.3)$, $r \sim U(0.04, 0.08)$
    \item Target 7: $x \sim U(0.225, 0.35)$, $y \sim U(-0.15,0.15)$,
    
    $z \sim U(-0.3, 0.3)$, $r \sim U(0.05, 0.15)$
\end{itemize}
\subsubsection{Public Display Interaction}
The buttons are represented as vertical square boxes of face widths 5 cm with their $[y,z]$ coordinates in meters set to $[-0.15, 0.01]$, $[0.15, 0.01]$, $[-0.15, -0.29]$, and $[0.15, -0.29]$, on a flat vertical plane 42 cm in front of the user's shoulder, respectively. All of these buttons require a minimum touch force of 1 N to successfully register the button being pressed.
\subsubsection{Mobile Keyboard Typing}
The simulated user is tasked with pointing to five small targets located on a horizontal plane, resembling typing on a five-character keyboard on a mobile device. 
We use spheres of radius 1 cm to define the pointing targets, which have a dwell duration of $50$ ms. The spheres are randomly spawned on a horizontal plane of size $15$ cm $\times$  $7.2$ cm located 30 cm below and 30 cm in front of the user's shoulder.

\onecolumn
\section{Brax PPO Hyperparameters}
\label{sec:hyperparams}
\begin{table}[H]
\centering
\begin{tabular}{ll}
\hline
\textbf{Parameter} & \textbf{Value}\\
\hline
Number of Environments & 1024 \\
Batch Size & 128 \\
Number of Mini-batches & 8 \\
Number of Updates per Batch & 8 \\
Unroll Length & 10 \\
Entropy Cost & 0.001 \\
Discount Factor & 0.97 \\
Learning Rate & 0.0003 \\
Reward Scaling & 0.1 \\
Clipping Epsilon & 0.3 \\
Policy Network MLP Layer Sizes & [128, 128, 128, 128] \\
Value Network MLP Layer Sizes & [256, 256, 256, 256, 256] \\
\hline
\end{tabular}
\caption{Brax PPO hyperparameters used as defaults by our framework.}
\Description{Table of default Brax PPO hyperparameters showing: 1024 environments, batch size 128, 8 mini-batches, 8 updates per batch, unroll length 10, entropy cost 0.001, discount factor 0.97, learning rate 0.0003, reward scaling 0.1, clipping epsilon 0.3, a 4-layer policy network (128 units each), and a 5-layer value network (256 units each).}
\end{table}
\section{Biomechanical Validation for MyoInteract (Replication)}
\label{sec:validation_replication}
\begin{figure}[H]
    \centering
    \subfloat[\centering Fitts' Law Evaluation \label{fig:fitts_law_replication}]{\includegraphics[width=0.3\linewidth]{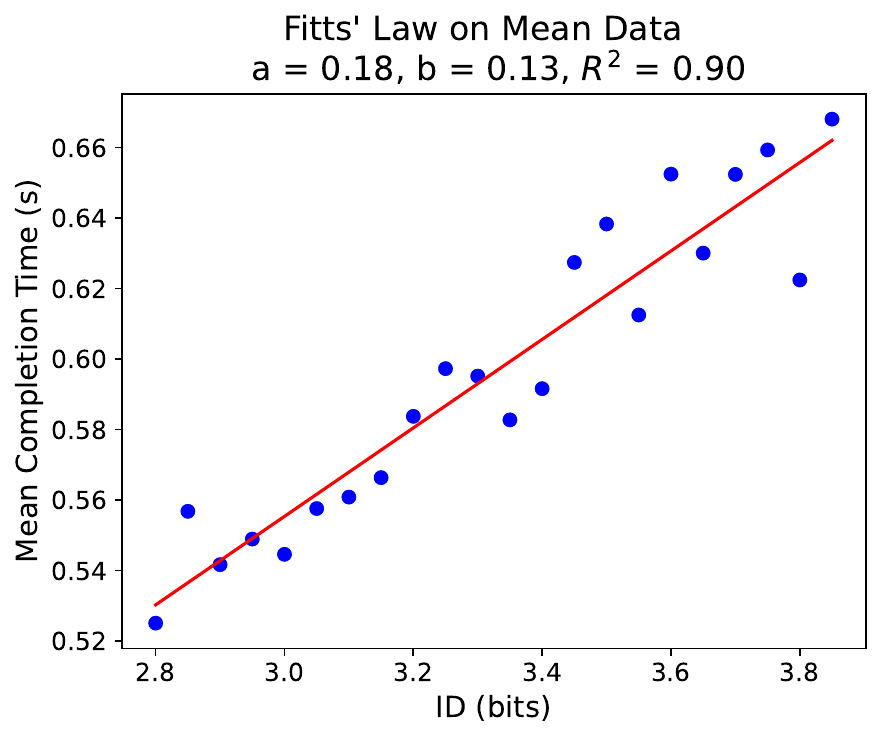}}
    \qquad
    \subfloat[\centering Projected Positions \label{fig:projected_positions_replication}]{{\includegraphics[width=0.3\linewidth]{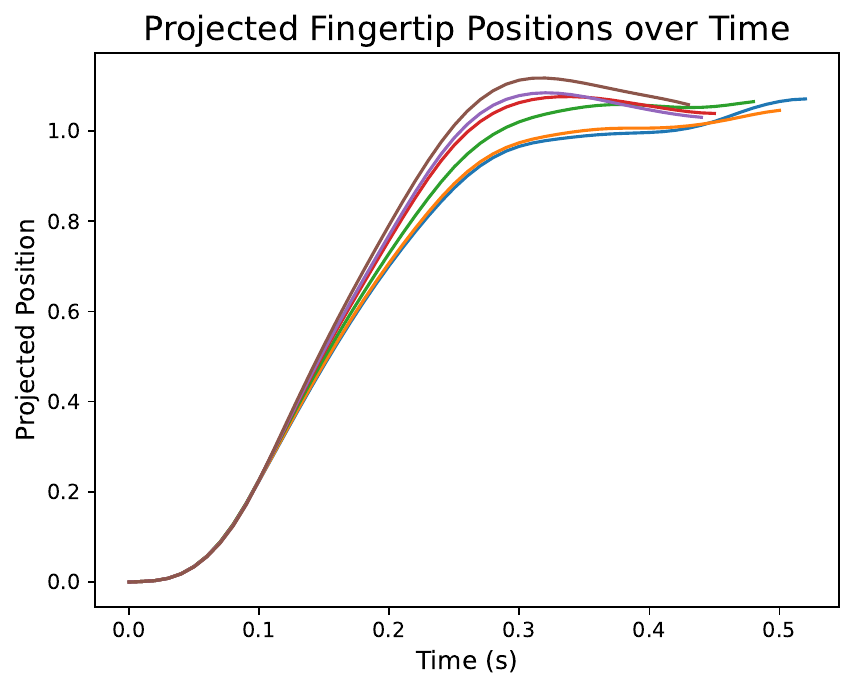} }}
    \qquad
    \subfloat[\centering Projected Velocities \label{fig:projected_velocities_replication}]{{\includegraphics[width=0.3\linewidth]{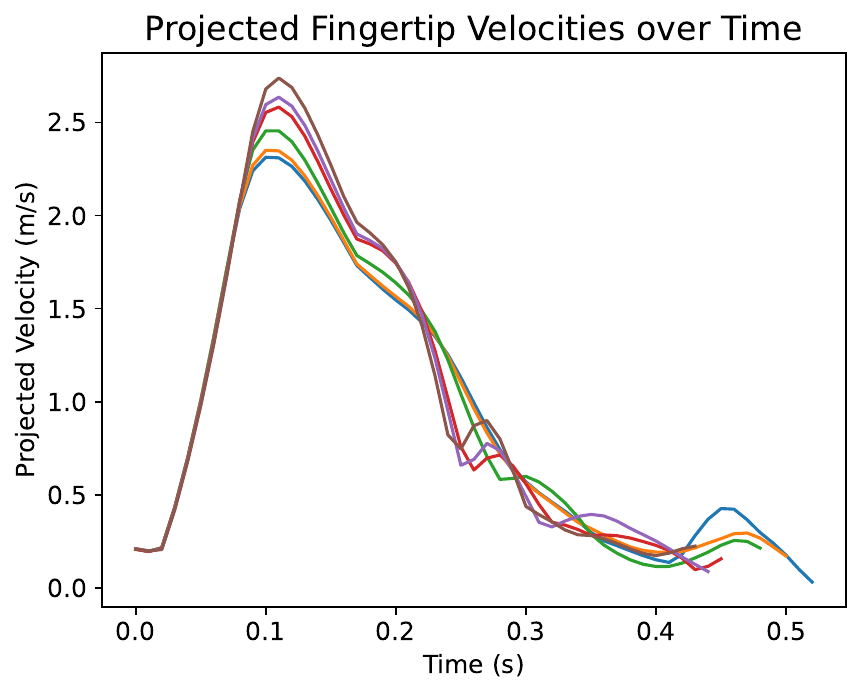} }}
    \caption{Biomechanical validation of the policy trained using \textbf{MyoInteract (Replication)}.\\
    (a) Average movement times (blue dots) plotted against IDs of targets. The observed movement times follow a clear linear relationship (red line), as predicted by Fitts' Law, with a fit of $R^2 = 0.90$.
    (b) Projected fingertip positions for a trained policy, moving from the initial position (0) to the target position (1) and settling there.
    (c) Corresponding projected fingertip velocities, exhibiting the characteristic bell-shaped profile of aimed reaching movements~\cite{morasso1981spatial}.}
    \label{fig:projected_pos_vel_replication}
    \Description{Validation of a trained policy using biomechanical evaluation, consisting of three plots. The first plot, labeled "Fitts' Law Evaluation," shows a scatter plot of average movement times against target IDs, with a linear line illustrating fitted model predictions (red line) and a fit of R2=0.90, indicating that the observed movement times follow Fitts' Law.
    The second plot, "Projected Fingertip Positions," displays the projected fingertip positions over time, illustrating a smooth movement from the initial position (0) to the target position (1) and stabilizing at the goal.
    The third plot, "Projected Fingertip Velocities," shows the corresponding projected fingertip velocities, exhibiting a characteristic bell-shaped profile of aimed reaching movements.}
\end{figure}
\clearpage
\section{Study Questions}\label{sec:study-questions}
\begin{table}[h!]
\centering
\caption{Overview of questionnaire items. The scales for questions 2 to 11 range from 1 (strongly disagree) to 5 (strongly agree). For questions 16-18, the user selects between `Novice', `Intermediate', and `Expert'.}
\label{table:study-questions}
\begin{tabular}{@{}p{0.5cm}p{13cm}p{0.75cm}@{}}
\toprule
\textbf{No.} & \textbf{Question}                                                                                                                                               & \textbf{Type} \\ \midrule
             & \multicolumn{2}{l}{\textbf{Policy Evaluation}}                                                                                                                                  \\
1            & Please describe whether the trained policy satisfied your expectations. If the policy was not successful, how would you change the parameters in the next step? & text          \\
             & \multicolumn{2}{l}{\textbf{RL Setup and Training}}                                                                                                                              \\
2            & It was easy for me to define and set up a Reinforcement Learning task.                                                                                          & scale         \\
3            & It was difficult for me to define and set up a Reinforcement Learning task.                                                                                     & scale         \\
4            & It was difficult for me to train a simulated user.                                                                                                              & scale         \\
5            & It was easy for me to train a simulated user.                                                                                                                   & scale         \\
6            & Which aspects of the UI did you find useful for (1) defining a Reinforcement Learning task and (2) training a simulated user?                                   & text          \\
7            & What could be improved in the UI to simplify (1) defining a Reinforcement Learning task and (2) training a simulated user?                                      & text          \\
             & \multicolumn{2}{l}{\textbf{Policy Monitoring and Analysis}}                                                                                                                     \\
8            & It was difficult for me to monitor the progress of the training.                                                                                                & scale         \\
9            & It was easy for me to monitor the progress of the training.                                                                                                     & scale         \\
10           & It was difficult for me to assess the quality of the trained policy.                                                                                            & scale         \\
11           & It was easy for me to assess the quality of the trained policy.                                                                                                 & scale         \\
12           & Which metrics and logs did you find useful to (1) monitor the training progress and (2) assess a learned policy?                                                & text          \\
13           & Which other metrics and logs would you find useful to (1) monitor the training progress and (2) assess a learned policy?                                        & text          \\
             & \multicolumn{2}{l}{\textbf{Demographics}}                                                                                                                                       \\
14           & Age                                                                                                                                                             & text          \\
15           & Gender                                                                                                                                                          & text          \\
16           & What is your experience level with Reinforcement Learning?                                                                                                      & scale         \\
17           & What is your experience level with Biomechanical Models?                                                                                                        & scale         \\
18           & What is your experience level with Cognitive Models?                                                                                                            & scale         \\ \bottomrule
\end{tabular}
\Description{Overview of 18 questionnaire items across four categories. "Policy Evaluation" includes one open-text question. "RL Setup and Training" includes four Likert-scale items (questions 2–5) on ease and difficulty of defining RL tasks and training simulated users, plus two open-text questions. "Policy Monitoring and Analysis" includes four Likert-scale items (questions 8–11) on ease and difficulty of monitoring training and assessing policy quality, plus two open-text questions. "Demographics" collects age, gender, and three experience-level ratings (Reinforcement Learning, Biomechanical Models, Cognitive Models) on a novice/intermediate/expert scale.}
\end{table}

\clearpage
\section{GUI}
\begin{figure}[!ht]
    \centering
    \includegraphics[width=0.9\linewidth]{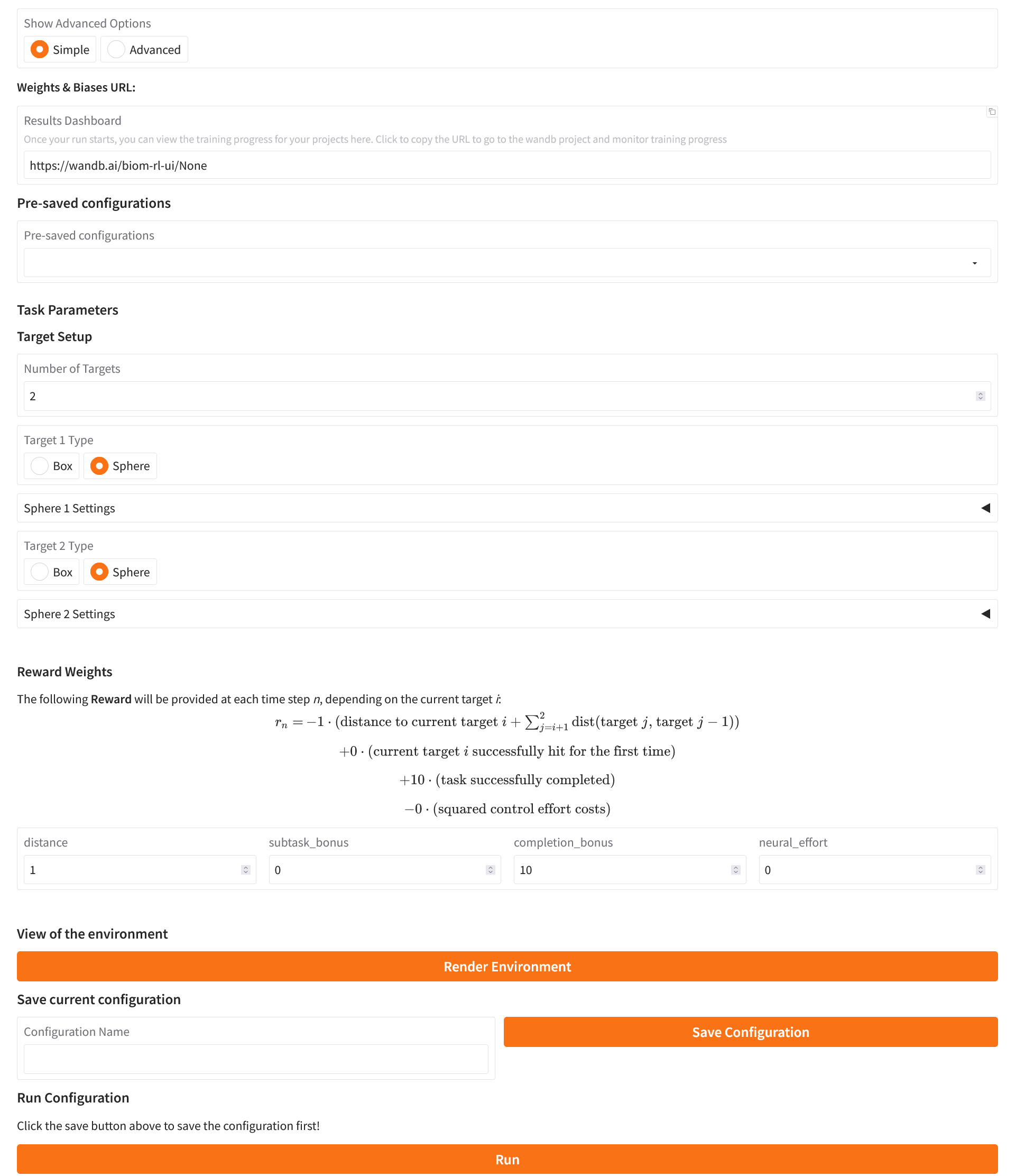}
    \caption{Simple version of our GUI for configuring reinforcement learning experiments, adjusting task and reward parameters, rendering the environment, and running training sessions.}
    \label{fig:simple_GUI}
    \Description{The simple mode of the graphical user interface (GUI) for configuring reinforcement learning experiments. The GUI allows users to adjust various parameters, including task parameters, target setup, and reward weights. The task parameters section includes options for setting the number of targets, target types (box or sphere), and target settings. The reward weights section provides a mathematical formula for calculating rewards at each time step, with adjustable weights for distance, subtask bonus, completion bonus, and neural effort. The GUI also features options for rendering the environment, saving configurations, and running training sessions.}
\end{figure}
\begin{figure}[!ht]
    \centering
    \includegraphics[width=0.9\linewidth]{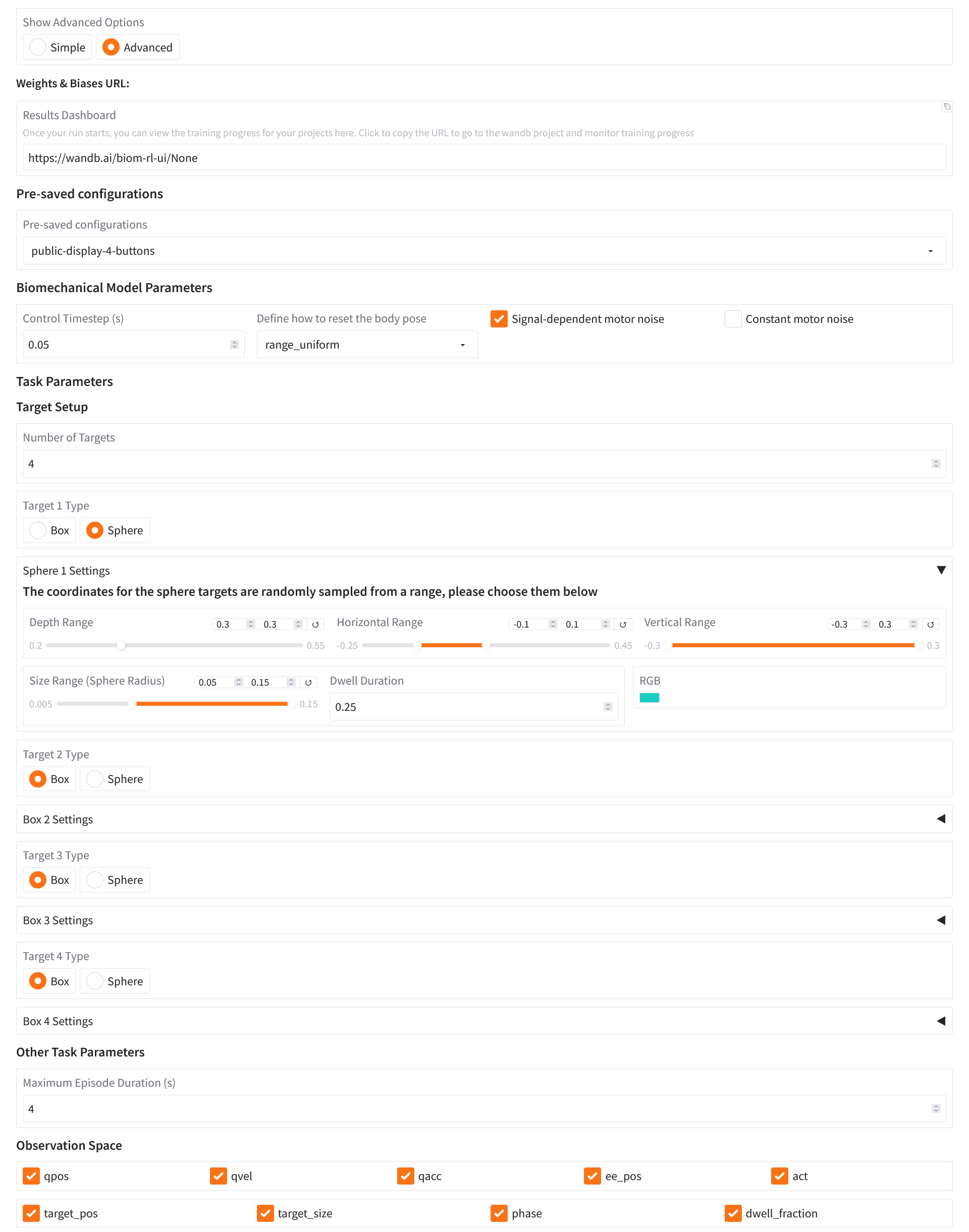}
    \caption{Advanced version of our GUI for configuring reinforcement learning experiments, adjusting task and reward parameters, rendering the environment, and running training sessions.
    In addition to the settings present in Simple Mode, this version enables modifying biomechanical properties, observation space components RL hyperparameters, and initialising the training with previously saved checkpoints
    (continued on next page).}
    \Description{The first part of the advanced mode of the graphical user interface (GUI) for configuring reinforcement learning experiments. The GUI provides various options for adjusting task and reward parameters, rendering the environment, and running training sessions. The interface includes sections for Biomechanical Model Parameters, allowing users to adjust control time step, body pose reset, motor noise, and other parameters. The Task Parameters section enables users to configure target setup, including number of targets, target types, and target settings. There is an input field for the maximum episode duration and boxes that can be clicked to be included in the observation space (qpos, qvel, qacc, ee_pos, act, target_pos, target_size, phase, dwell_fraction).}
    \label{fig:advanced_GUI}
\end{figure}
\begin{figure}[!ht]
    \ContinuedFloat
    \centering
    \includegraphics[width=0.9\linewidth]{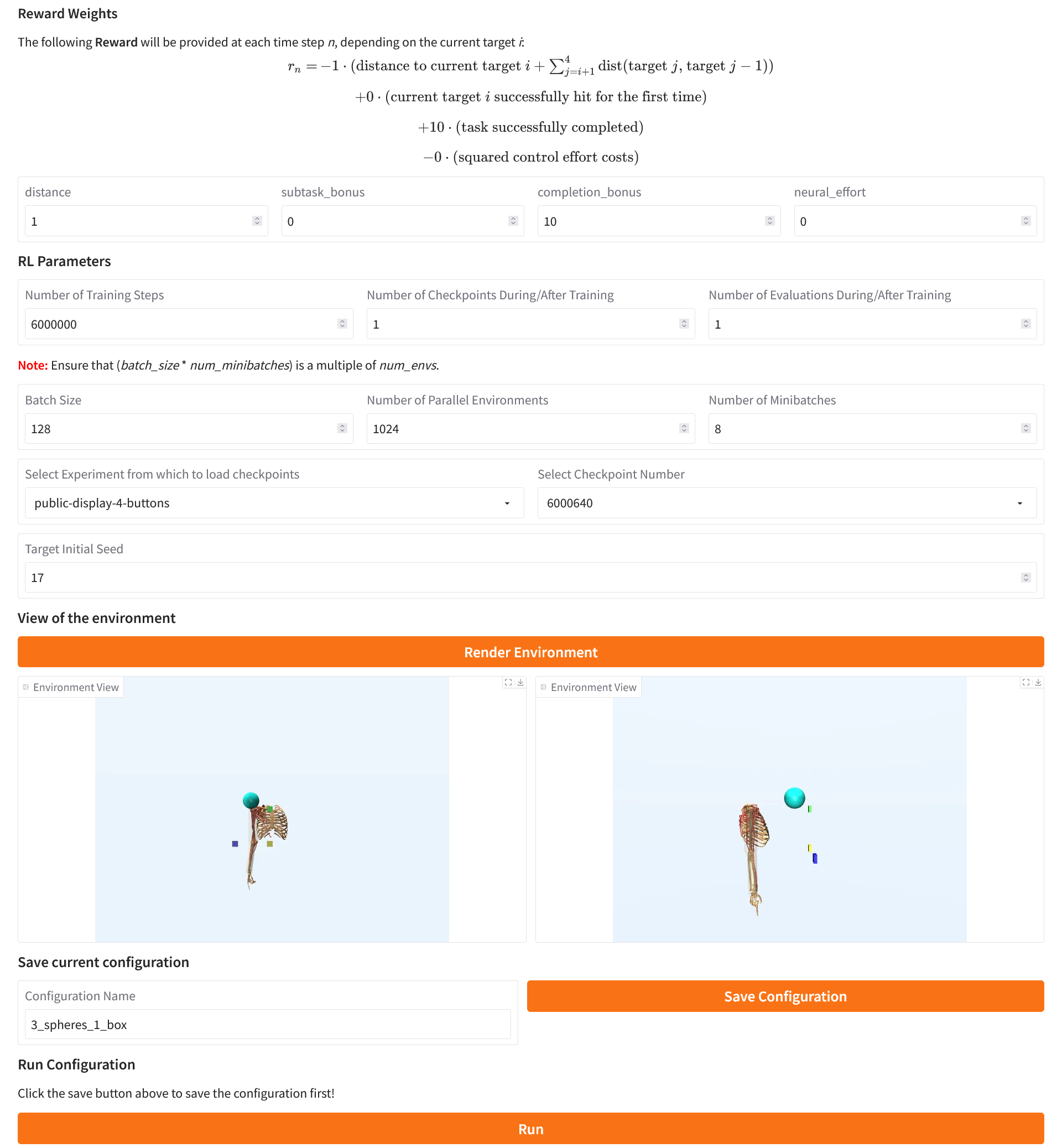}
    \caption[]{Advanced version of our GUI (continued from previous page).}
    
    \Description{The second part of the advanced mode of the graphical user interface (GUI) for configuring reinforcement learning experiments. The Reward Weights section provides a mathematical formula for calculating rewards at each time step, with adjustable weights for distance, subtask bonus, completion bonus, and neural effort. The RL Parameters section allows users to adjust RL parameters, such as number of training steps, number of checkpoints during/after training, number of evaluations during/after training, batch size, number of parallel environments, and number of minibatches. 
    Then there is a dropdown to choose an already trained checkpoint to start from and an initial seed. The GUI also features options for rendering the environment, saving configurations, and running training sessions.}
\end{figure}

\end{document}